\documentclass[11pt]{article}
\usepackage{amsmath,amsfonts,setspace,cancel,url,hyperref,verbatim,soul,cite}

\usepackage{fullpage}
\usepackage{slashed}
\usepackage{cancel}
\newcommand{\be}{\begin{equation}}
\newcommand{\ee}{\end{equation}}
\newcommand{\gM}{\mathcal{H}}

\newcommand{\vol}{\mathrm{Vol}(S^7)}
\numberwithin{equation}{section}

\begin{document}
\begin{titlepage}
\begin{center}

\hfill  DAMTP-2016-55

\vskip 1.5cm

{\Large \bf  Black hole thermodynamics, stringy dualities and double field theory}

\vskip 2cm

{\bf Alex S.~Arvanitakis\,${}^1$  and 
Chris D.~A.~Blair\,${}^2$} \\

\vskip 15pt

{\em $^1$ \hskip -.1truecm
\em  Department of Applied Mathematics and Theoretical Physics,\\ Centre for Mathematical Sciences, University of Cambridge,\\
Wilberforce Road, Cambridge, CB3 0WA, U.K.\vskip 5pt }

{email: {\tt A.S.Arvanitakis@damtp.cam.ac.uk}} \\

\vskip .4truecm

{\em $^2$ \hskip -.1truecm
\em Theoretische Natuurkunde, Vrije Universiteit Brussel, and the International Solvay Institutes, \\ 
            Pleinlaan 2, B-1050 Brussels, Belgium \\ 
\vskip 5pt }

{email: {\tt cblair@vub.ac.be}} \\
\end{center}

\begin{abstract}

We discuss black hole thermodynamics in the manifestly duality invariant formalism of double field theory (DFT).
We reformulate and prove the first law of black hole thermodynamics in DFT, using the covariant phase space approach.
After splitting the full $O(D,D)$ invariant DFT into a Kaluza-Klein-inspired form where only $n$ coordinates are doubled, our results provide explicit duality invariant mass and entropy formulas.
We illustrate how this works by discussing the black string solution and its T-duals.

\end{abstract}
\end{titlepage}

\tableofcontents
\section{Introduction}

The massless spectrum of any of the closed string theories has a common sector consisting of the NSNS fields: the spacetime metric $g_{\mu\nu}$, 2-form $B_{\mu\nu}$ and dilaton $\phi$. The low-energy effective action for these fields is: \cite{Callan:1985ia} 
\be
\label{spacetime_action}
S_{NSNS} =  \int d^{D}x \: \sqrt{\det g} e^{-2\phi} \left(R + 4 (\nabla\phi)^2 -\frac{1}{12} H^2 \right)\,, \qquad H_{\mu\nu\rho} \equiv  3\partial_{[\mu}B_{\nu\rho]}\,.
\ee
A solution of the equations of motion with an isometry direction $z$ can be mapped into another solution by the Buscher rules \cite{Buscher:1987qj,Buscher:1987sk}. In terms of a coordinate split $x^\mu = (x^i,z)$, these rules are:
\begin{align}
\label{Buscher}
\tilde g_{zz}=g_{zz}^{-1}\,,\quad \tilde g_{z i}=g_{zz}^{-1} B_{z i} \,,\quad\tilde B_{z i} = g_{zz}^{-1} g_{z i} \,,\quad e^{-2\tilde \phi} \sqrt{\det \tilde g}= e^{-2\phi} \sqrt{\det g}\,,\nonumber\\
\tilde g_{ij}=g_{ij}- g_{zz}^{-1}(g_{zi}g_{zj} - B_{zi} B_{zj})\,,\quad \tilde B_{ij}= B_{ij} - g_{zz}^{-1}(g_{zi} B_{z j} - B_{z i} g_{zj}) \,.
\end{align}
Under certain conditions, e.g. when $z$ is a compact isometry direction \cite{Rocek:1991ps}, the pair of solutions are \emph{equivalent} in the sense e.g. that the string sigma-models defined on either solution define the same CFT. In the presence of $n$ isometries, the possible duality transformations form the group $O(n,n)$. 

A recent development (which however has earlier roots in \cite{Duff:1989tf, Siegel:1993th, Siegel:1993xq}) is double field theory (DFT) \cite{Hull:2009mi, Hull:2009zb, Hohm:2010jy, Hohm:2010pp}, which aims to describe supergravity in a manifestly duality invariant manner. In its most conservative interpretation double field theory can be viewed as a reorganisation of the degrees of freedom $(g,B,\phi)$ of \eqref{spacetime_action} into objects that transform \emph{linearly} under the T-duality group $O(D,D)$. These objects consist of a generalised metric, $H_{MN}$, which is a rank 2 tensor under T-duality transformations, and the generalised dilaton, $d$, which is invariant. 

Double field theory involves the introduction of dual coordinates $\tilde x_\mu$ such that all fields and gauge parameters may depend in principle on the $2D$ coordinates $X^M = ( x^\mu, \tilde x_\mu)$, which form an $O(D,D)$ vector. The local symmetry transformations of \eqref{spacetime_action} --- diffeomorphisms and gauge transformations of the 2-form --- are replaced in DFT by so-called \emph{generalised diffeomorphisms}. These provide infinitesimal $O(D,D)$ transformations, just as diffeomorphisms in general relativity provide an action of $\mathrm{GL}(D)$.

This doubling leads to constraints, not unexpectedly. In order for the action of generalised diffeomorphisms to give a closed algebra, one is forced to impose conditions on the fields and gauge parameters of DFT. The simplest choice is simply to require that we can only depend on at most half of the coordinates, in which case the DFT action --- which is fixed by invariance under generalised diffeomorphisms, modulo the closure constraints --- can be reduced back to that of \eqref{spacetime_action}. This constraint goes by the name of the \emph{section condition} or \emph{strong constraint}. However, one can also achieve closure by requiring a generalised Scherk-Schwarz factorisation of fields and gauge parameters, in which case dependence on dual coordinates through twist matrices is possible \cite{Aldazabal:2013sca}. This leads to gauged supergravities whose higher-dimensional origins were previously unknown. 

This suggests that double field theory provides a framework to study possibilities suggested by T-duality which go beyond supergravity. 
It is also believed to be the natural setting for the description of ``non-geometric'' backgrounds, such as exotic branes \cite{deBoer:2012ma}, where the spacetime fields are patched together by duality transformations.
Indeed, this was one of the original motivations for the development of the theory. 

A major goal for the theory is therefore to study the form and properties of double field theory backgrounds.

Of course, imposing the section condition implies that all supergravity backgrounds can be viewed as solutions of double field theory. The interpretation of solutions in the doubled space is still of interest. The standard 1/2-BPS solutions of the NSNS sector are the fundamental string (F1) and its T-dual, the pp-wave, plus their magnetic counterparts, the NS5 brane and the Kaluza-Klein monopole (KKM). These have been investigated as double field theory solutions \cite{Berkeley:2014nza, Berman:2014jsa, Berman:2014hna}, revealing that one can think of such solutions as simply waves or monopoles embedded in the doubled space, with for instance the orientation of the wave relative to the choice of section determining whether the solution appears in spacetime as a string or a wave. A recent extension of this approach to study non-geometric branes as DFT solutions was considered in \cite{Bakhmatov:2016kfn}.\footnote{Ideally, of course, one would like to construct genuinely doubled backgrounds which would not be 
admissible in standard supergravity. See for instance the discussions in \cite{Berman:2014jsa, Bakhmatov:2016kfn} concerning the possibility of having solutions with depend on $\tilde x$ --- a dual coordinate --- but not $x$: these obey the section condition but are certainly non-geometric in the ``physical'' frame where $x$ is a coordinate.}

In supergravity, one can construct notions of mass (via the ADM formula) and charge (via integrals of field strengths and their duals). In double field theory, both gravity and the $B$-field appear together in the generalised metric. In \cite{Blair:2015eba, Park:2015bza}, it was shown that the appropriate notion of conserved charges in DFT follows from applying a Noether procedure to the invariance of the DFT action under generalised diffeomorphisms: the electric charge of the $B$-field can be associated to translational invariance in a dual direction. Similar expressions were found using a Hamiltonian decomposition of the DFT action in \cite{Naseer:2015fba}. 

This leads to a nice understanding of the properties of 1/2-BPS branes within double field theory. It would be interesting to pursue the properties of backgrounds in DFT beyond this sector. In this paper, we intend to focus on \emph{non-extremal solutions}.\footnote{We note that the paper \cite{Park:2015bza} evaluated the charges of some black hole solutions in the context of the DFT current, finding the conventional supergravity results, however without studying their behaviour under T-duality.}
The study of black hole or black brane solutions in DFT should be interesting from a number of perspectives. We might wonder whether access to T-dual descriptions has implications for the notions of singularities and horizons. We are also interested in the description of thermodynamics. 

We know that various thermodynamic quantities (mass, entropy, \dots) associated to black holes should be (and, empirically, are) duality-invariant (see e.g. \cite{Mohaupt:2000mj} for a review). For black hole entropy, for example, this is intuitively obvious if geometries related by duality are supposed to provide equivalent descriptions of the underlying microscopic degrees of freedom (whatever those are), and if the entropy is supposed to provide a measure of the number of said degrees of freedom. 

The issue of duality invariance of entropy and other thermodynamic quantities has been looked into from a semi-classical gravity (or, macroscopic) perspective in a few works, of which \cite{Horowitz:1993wt} by Horowitz and Welch appears to be the earliest. They verify the invariance of the surface gravity and horizon area of a black hole with bifurcate Killing horizon under a Buscher transformation \eqref{Buscher} by an explicit 
component calculation in spacetime.

It seems more natural, however, to examine duality-invariant properties in a formalism where duality invariance is \emph{manifest} from the outset. In our work we therefore investigate black hole thermodynamics in DFT. 

The main result of our investigation is the duality-invariant black hole mass and entropy formulas \eqref{massdef} and \eqref{entropy_def} satisfying the first law of black hole thermodynamics \eqref{1stlawintegralfinal} (where the invariance is in fact under the subgroup $O(n,n)$ for $n \leq D-2$). Momentum, angular momentum and winding charge all enter the first law in manifestly duality-invariant combinations. 

To derive this result, we make use of the ``covariant phase space'' approach due to Lee, Iyer, and Wald \cite{Lee:1990nz,Wald:1993nt,Iyer:1994ys}. In this approach, the first law of black hole thermodynamics is re-expressed in a ``differential'' form, as the vanishing of the exterior derivative of a certain $(D-2)$-form constructed out of the fields and their variations; Stokes' theorem then sets the integral of this form on a horizon cross-section (which is related to the variation of the entropy) equal to the integral on a sphere at infinity (which yields variations of energy, angular momentum, etc.), recovering the usual, integrated form of the first law \eqref{1stlawintegralfinal}. 

Although DFT does not admit the standard notion of a differential form, one can work instead with 
contravariant antisymmetric $O(D,D)$ tensor densities (as detailed in appendix \ref{appendix_integration}), and one can express the first law of black hole thermodynamics in differential form as
\be
\partial_P\delta \mathcal{Q}^{MP}=0\,,
\ee
where $\delta \mathcal{Q}^{MP}$ is an expression \eqref{charge_integrand} constructed out of the generalised metric $H_{MN}$, generalised dilaton $d$, and their variations {in a fully $O(D,D)$ covariant fashion. 

In the next section we provide a brief introduction to double field theory and also clarify certain subtleties which will be relevant later. In section \ref{DFTfirstlaw} we present a derivation of the first law of black hole thermodynamics in DFT through the covariant phase space formalism. After providing a brief motivating example from particle mechanics, we proceed to use the formalism to derive expressions for the Noether charge associated to a generalised Killing vector $\Lambda^M$, and prove the first law in its ``differential'' form \eqref{1stlawdifferential}. Then in section \ref{split}  we partially break $O(D,D)$ to $O(n,n)$ $(n\leq D-2)$ using the split parametrisation \eqref{KK_parametrisation} in order to decompose said Noether charge into entropy, mass, momentum and winding charges and show that the variations of these charges satisfy the first law of black hole thermodynamics \eqref{1stlawintegralfinal}. Section \ref{example} is devoted to an analysis of the black string solution of \eqref{
spacetime_action} from the DFT point of 
view and a verification of our mass and entropy formulas. We conclude with a discussion of our results and possible generalisations thereof. We also provide appendices containing additional results, including a discussion of Stokes' theorem in DFT.

\section{Double field theory}
\label{two} 

\subsection{Double field theory in a nutshell}

The group $O(D,D)$ itself is defined to be the set of transformations preserving the $O(D,D)$ structure: 
\be
\eta_{MN} = \begin{pmatrix} 0 & \mathbb{I}_D \\ \mathbb{I}_D & 0 \end{pmatrix} \,,
\ee
which will be used to raise and lower indices below. The fields of double field theory in the NSNS sector are the generalised metric, $H_{MN}$, and the generalised dilaton, $d$. The generalised metric is symmetric and constrained to satisfy $H_{M}{}^N H_{N}{}^P = \delta_M{}^P$, which implies that it parametrises the coset $O(D,D)/ O(1,D-1) \times O(1,D-1)$. 

The double field theory action is 
\be
\label{dftaction}
S_\text{DFT}= \frac{1}{16\pi G_{DFT}} \int d^{2 D} X \: e^{-2d} \mathcal{R}
\ee
where the generalised Ricci scalar $\mathcal{R}$ is
\be
 \begin{split}
  \mathcal{R} =  \ &\;4\,H^{MN}\partial_{M}\partial_{N}d
  -\partial_{M}\partial_{N}H^{MN} 
    -4\,H^{MN}\partial_{M}d\,\partial_{N}d
   + 4 \partial_M H^{MN}  \,\partial_Nd\;\\[1.0ex]
    ~&+\frac{1}{8}\,H^{MN}\partial_{M}H^{KL}\,
  \partial_{N}H_{KL}-\frac{1}{2}H^{MN}\partial_{M}H^{KL}\,
  \partial_{K}H_{NL}\;.
 \end{split}
\ee
We will define $G_{DFT}$ below.

The action is fixed by requiring invariance under generalised diffeomorphisms. 
These are parametrised by a generalised vector, $\Lambda^M$, and act on the fields through a generalised Lie derivative, denoted $\mathcal{L}_\Lambda$, such that on a generalised vector $V^M$ we have
\be
\delta_\Lambda V^M  \equiv  \mathcal{L}_\Lambda V^M = \Lambda^N \partial_N V^M - V^N \partial_N \Lambda^M + \partial^M \Lambda_N V^N\,.
\label{gendiffeodef} 
\ee
By construction, this generalised Lie derivative preserves the $O(D,D)$ structure $\eta_{MN}$. 

The generalised dilaton transforms such that $e^{-2d}$ is a scalar of weight 1 (and thus provides a measure for integration), while the generalised metric $H_{MN}$ transforms as a symmetric rank 2 tensor, as indicated  by its pair of $O(D,D)$ indices. 
\be
\label{eq:gendiffeo}
\delta_\Lambda H_{MN} =\Lambda^P\partial_P H_{MN}+ 2(\partial_{(M} \Lambda^P - \partial^P \Lambda_{(M}) H_{N) P}  \,,\qquad \delta_\Lambda e^{-2d} = \partial_{P}\left( e^{-2d } \Lambda^P\right)\,.
\ee
As we mentioned in the introduction, closure of the algebra of generalised diffeomorphisms leads to constraints. 
The closure condition is 
\be
\mathcal{L}_{\Lambda_1} \mathcal{L}_{\Lambda_2} - \mathcal{L}_{\Lambda_2} \mathcal{L}_{\Lambda_1} 
= \mathcal{L}_{ [ \Lambda_1, \Lambda_2 ]_C }  \,,
\ee
where the antisymmetric bracket (generalising the Lie bracket) is 
\be
[ \Lambda_1, \Lambda_2 ]_C = \frac{1}{2} ( \mathcal{L}_{\Lambda_1} \Lambda_2 - \mathcal{L}_{\Lambda_2} \Lambda_1 )\,.
\label{courant} 
\ee
Closure can be guaranteed by requiring the section condition: 
\begin{align}
\label{strongconstraint}
\partial_M A  \partial^M B =0 
\quad , \quad \partial_M \partial^M A = 0 \,, 
\end{align}
acting on $A,B$ any fields and gauge parameters in the theory. 
This constraint is solved (locally) by assuming all fields and gauge parameters depend on half the coordinates in $X^M$; this is called choosing a \emph{section}. Choosing a section breaks $O(D,D)$ invariance, but the strong constraint itself is an invariant statement. The usual notion of T-duality is recovered if after choosing the section there remain some number of isometries: these give an ambiguity in the choice of section corresponding to different duality frames. 

If we write $X^M=(x^\mu, \tilde x_\mu)$ and choose the section $\tilde \partial^\mu =0$, then parametrising the DFT fields $(d,H)$ in terms of the spacetime fields $(g,B,\phi)$ as
\be
\label{dft_spacetime_parametrisation}
H_{MN}=\begin{pmatrix} g_{\mu\nu} - B_{\mu\rho}g^{\rho \sigma} B_{\sigma \nu} & B_{\mu\rho} g^{\rho \nu} \\ -g^{\mu\rho} B_{\rho \nu} & g^{\mu\nu}\end{pmatrix}\,, \qquad e^{-2d}= e^{-2(\phi-\phi_0)}\sqrt{\det g}
\ee 
one finds that the DFT action \eqref{dftaction} reduces to 
\be
\label{vev-subtracted-action}
S_{NSNS} = \frac{ e^{2\phi_0} }{ 16 \pi G_D }\int d^Dx \: \sqrt{|g|} e^{-2\phi} \left(R + 4(\nabla \phi)^2 -\frac{1}{12} H^2 \right) \,,
\ee
where in $D$ dimensions $G_D \propto l_s^{D-2} e^{2\phi_0}$, in particular in $D=10$ we have the usual constant 
\be
G  \equiv  G_{10} = 8 \pi^6 l_s^8 e^{2 \phi_0} \,;
\ee
this means that we have defined
\be
G_{DFT} \equiv G_D \int d^D\tilde x \,.
\label{eq:GDFT}
\ee
In general, one may think of this as a formal expression designed to cancel the integration over the dual coordinates. In the case where we are dealing with a doubled torus, with physical radii $R_i$ and dual radii $\tilde R_i = l_s^2/R_i$, we have the explicit T-duality invariant form 
\be
\begin{split} 
G_{DFT} & = (2\pi)^D \tilde R_1 \dots \tilde R_D G_D \\ 
& = (2\pi)^D  R_1 \dots R_D \tilde G_D\,, \\ 
\end{split} 
\ee
which, given that $G_D \propto e^{2\phi_0}$, is only consistent if the dilaton transforms so that
\be
e^{2\tilde \phi_0}  = e^{2\phi_0} \frac{ (l_s)^{2D}}{(R_1 \dots R_D)^2}\,,
\ee
which is the correct transformation rule when dualising in $D$ dimensions. Note that we are taking our coordinates here to have the range $[0, 2\pi R]$ so that the information about their radii is contained here and not in the metric (i.e. we will write expressions like $ds^2 = g_{xx} dx^2$ and implement the Buscher rule simply as $g_{xx} \leftrightarrow 1/ g_{xx}$ consistent with the form \eqref{dft_spacetime_parametrisation} of the generalised metric, so that there are no hidden $l_s^2$). This accounts for the appearance of the asymptotic value of the dilaton in \eqref{dft_spacetime_parametrisation}, which we have included to take into account the transformation $e^{2\tilde\phi_0} = e^{2\phi} l_s^2/R^2$ which we would otherwise miss. It is important to have the correct prefactors in place to correctly measure charges.  

The above definition \eqref{eq:GDFT} corresponds essentially to the discussion in \cite{Berman:2014hna}. Here we have attempted to be a little bit more precise, especially concerning the dilaton.

\subsection{On curvature, singularities and horizons} 

The geometry of double field theory is based on generalised diffeomorphisms, as defined in \eqref{gendiffeodef}, and hence is not that of conventional differential geometry \cite{Siegel:1993xq, Siegel:1993th, Jeon:2010rw, Jeon:2011cn,Hohm:2011si, Berman:2013uda}. Thus a connection in DFT provides a covariant derivative which is covariant under generalised diffeomorphisms. One can define for a connection a generalised Riemann tensor, $\mathcal{R}_{MNP}{}^Q$, and a generalised torsion $\tau_{MN}{}^P$, which do not coincide with the usual definitions.

The natural connection in Riemannian geometry is the Levi-Civita connection. 
In DFT, one would analogously seek to produce a connection compatible with both the generalised metric and the $O(D,D)$ structure, with vanishing generalised torsion (and also compatible with using $e^{-2d}$ as the integration measure). 

These conditions do not have a unique solution. The connection coefficients can only be found up to some number of components which cannot be determined in terms of the physical fields. These components can be projected out, using the projectors defined by $(P_\pm)_M^N = \frac{1}{2} ( \delta_M^N \pm H_M{}^N )$, so that covariant derivatives of tensors can still be well-defined if appropriately projected \cite{Siegel:1993th, Hohm:2011si}. 
(A ``dual'' point of view is to effectively set these undetermined components equal to zero, resulting in a so-called ``semi-covariant'' connection \cite{Jeon:2010rw, Jeon:2011cn}. Although setting the undetermined components equal to zero is certainly not a covariant condition, one can still construct covariant derivatives by projecting away the non-covariant transformations -- hence the name.)

The generalised Riemann tensor of such a connection has undetermined components (or is at best semi-covariant). Again, one can use the projectors to ameliorate the situation somewhat: the generalised Ricci tensor, $\mathcal{R}_{MN}$, and scalar, $\mathcal{R}$, can be defined by first projecting the generalised Riemann tensor and then contracting, producing expressions which \emph{are} uniquely determined in terms of the physical fields. 

These same expressions in fact follow also as the equations of motion of the generalised metric and dilaton:
\be
\mathcal{R} = 0 \quad , \quad
\mathcal{R}_{MN} = 0 \,.
\ee
We see therefore that the only completely physical and covariant curvature-like expressions in DFT vanish automatically by the equations of motion, at least away from sources.
As such, it seems that there is no way to measure curvature -- and hence curvature singularities -- in DFT. (If we we include the RR sector and fermions, then generically $\mathcal{R} \neq 0$, $\mathcal{R}_{MN} \neq 0$. But if they do not provide a good notion of curvature in the pure NSNS sector, then there is no reason to think they will do so then.)

There are also difficulties with higher-order curvature invariants. For instance, in \cite{Hohm:2011si} it was shown that there exists no scalar quantity in DFT which reduces to give the square of the Riemann tensor in spacetime. Such higher-order curvature terms appear of course as $O(\alpha^\prime)$ corrections to supergravity, and can be accomodated in DFT at $O(\alpha^\prime)$ through non-covariant field redefinitions leading to deformed gauge transformatons \cite{Hohm:2011si,Hohm:2014xsa, Marques:2015vua}. As we are interested in exploring properties of solutions to the theory at zeroth order in $\alpha^\prime$, we cannot access such higher order terms. 

In this paper, we will be interested in charges defined on Killing horizons of black hole solutions. A Killing horizon is a null hypersurface invariant under the action of a Killing vector $\xi$, on which the norm $\xi^2  =  g_{ij} \xi^i \xi^j$ vanishes. Although we know that the presence of horizon is preserved under Buscher duality along a spacelike symmetry \cite{Horowitz:1993wt}, under a duality along a \emph{timelike} duality, this is not so \cite{Rocek:1991ps}. In particular, for the Killing vector $\xi = \partial/\partial t$ present for static black holes, one knows that the Buscher rules involve inverting $g_{tt}  =  \xi^2$ which goes to zero at the horizon: hence in the dual solution, the horizon has been exchanged for a naked singularity. 

The full $O(D,D)$ formalism (for $D$ equal to the number of dimensions of spacetime) involves (perhaps formally) doubling all directions and so by default includes such timelike dualities. Hence, at least formally in DFT we see that horizons should be dual to singularities, and we have already seen that there does not seem to be a clear notion of the latter. 

A full $O(D,D)$-compatible definition of a horizon will not be given in this paper. Ideally, such a definition would involve some generalised Killing vector, $\Lambda$. However, it is not clear how to covariantly specify $\Lambda$ such that for instance the natural norm $H_{MN}\Lambda^M \Lambda^N$ reduces to the spacetime norm and then vanishes on a horizon. One possible approach is to use the idea of ``twisted vectors'' \cite{Hull:2014mxa}, however this involves knowing the $B$-field on each patch of the doubled spacetime and does not seem entirely satisfactory. 

We stress though that our results in the subsequent section will certainly continue to apply if or when a definition of a generalised Killing horizon is constructed, and they certainly make sense as they stand when one takes the point of view that they are valid for DFT backgrounds such that on some physical section there is the conventional notion of a horizon in spacetime. 

An alternative way out that sidesteps the issues of timelike dualities is to avoid doubling all directions, by making use of the ``Kaluza-Klein inspired'' split parameterisation of \cite{Hohm:2013nja}. Keeping in mind that the split parametrisation is equivalent to the usual, fully-doubled one, one can loosely think of this splitting as expressing the fully-doubled spacetime as a product of an ``external'', non-doubled geometry and an ``internal'' doubled geometry; one can then characterise a horizon lying purely within the external geometry in the usual way. Although the geometries we consider are not limited to such products, we will see in section \ref{split} how this strategy provides a definition of horizons which suffices for our purposes. This definition is also natural for the extension to EFT and the relationship to black holes of lower dimensonal SUGRA.

\section{Duality-invariant thermodynamics}
\label{DFTfirstlaw}

In this section, we shall derive the form of the first law of black hole thermodynamics in DFT, using the Lee-Iyer-Wald approach \cite{Lee:1990nz,Wald:1993nt,Iyer:1994ys} to conserved charges and black hole thermodynamics. These methods are appropriate for any diffeomorphism-invariant theory of gravity. Indeed, in \cite{Iyer:1994ys} Iyer and Wald provide a calculation of black hole entropy which goes through for any action where the gravitational degrees of freedom are encoded in the spacetime metric $g_{\mu\nu}$; this has become known as the Wald entropy formula.

Of course, in DFT we have not diffeomorphisms but generalised diffeomorphisms. However, as the arguments of Iyer and Wald fundamentally just require an action principle, they are straightforwardly adapted to DFT (this was already suggested in \cite{Park:2015bza}). 
In effect, as DFT unites the metric with the two-form, we are to some extent applying the method of \cite{Iyer:1994ys} to both of these fields simultaneously, and it is known that the latter can easily accommodate gauge fields (see e.g. the notes \cite{Compere:2006my}).

\subsection{The covariant phase space formalism}

Rather than launch directly into the full calculation in double field theory, we wish to first use this subsection to provide an introduction to the technology of the covariant phase space formalism by setting it in the simple and familiar context of Hamiltonian mechanics. 

Hamilton's equations for time evolution with Hamiltonian $H$ are
\be
0=\dot{x}^j \omega_{ij} - \partial_i H\,,
\label{HamEom} 
\ee
where $\omega_{ij}(x)$ is the symplectic form on a symplectic manifold (phase space) with coordinates $\{x^i\}$.

These equations can be derived from the action
\be
S= \int dt \: \left[ \dot x^i \theta_i(x) - H(x) \right]\,,
\ee
where $\theta_i(x)$ is a symplectic potential for $\omega_{ij}$\footnote{The symplectic form can be allowed to lie in a non-trivial cohomology class, which is to say that $\theta_i$ need not exist globally. The above action is perfectly well-defined regardless; see \cite{Howe:1989uk} for details. If $\omega_{ij}$ is indeed non-exact, the argument in this section implies that the Lee-Iyer-Wald symplectic form also fails to be exact, which is perhaps not obvious from its definition.}:
\be
\omega_{ij}(x)  =  2 \partial_{[i} \theta_{j]}(x)\,.
\ee
If we vary the action while keeping track of boundary terms we get
\begin{align}
S&=  \int dt \: \left[\frac{d}{dt}\left(\delta x^i \theta_i\right)\ +   \delta x^i \left( \dot{x}^j \omega_{ij} - \partial_i H\right)\right]\\
&=  \Theta[x;\delta x]+\int dt \: \left[\delta x^iE_i\right]\,,
\end{align}
where the equation of motion is exactly \eqref{HamEom}, $E_i \equiv \dot{x}^j \omega_{ij} - \partial_i H = 0$, and we have just defined the Lee-Iyer-Wald symplectic potential $\Theta[x;\delta x]$
\be
\Theta[x;\delta x] \equiv \int dt \:\frac{d}{dt}\left(\delta x^i \theta_i\right)\,.
\ee
In the Lee-Iyer-Wald covariant phase space formalism \cite{Lee:1990nz,Iyer:1994ys}, the symplectic form is defined as
\be
\Omega[x;\delta_1 x,\delta_2x] =  \delta_1 \Theta[x;\delta_2 x] -\delta_2 \Theta[x;\delta_1 x]\,.
\ee
For the above system we get
\be
\Omega[x;\delta_1 x,\delta_2x]= \int dt\: \frac{d}{dt}\left( \delta_1 x^i \delta_2 x^j \omega_{ij}\right)\,.
\ee
If we are considering an initial-value problem (which is standard in field theory, less so in particle mechanics), then the integral reduces to evaluation at initial time:
\be
\Omega[x;\delta_1 x,\delta_2x]=\delta_1 x^i \delta_2 x^j \omega_{ij}|_{t=0}\,.
\ee
We see that for an initial-value problem the Lee-Iyer-Wald symplectic form is identical to the standard one. Now if we define the functional $H[x]$
\be
H[x]  \equiv  \int dt \: H(x)
\ee
we can write down Hamilton's equations in the covariant phase space formalism:
\begin{align}
\forall \delta x\,,\quad \delta H[x] =\Omega[x;\delta x, \dot x]\,.
\end{align}
In particle mechanics, the above equation serves to identify the time evolution $\dot x$ on the right-hand side generated by $H$ on the left-hand side; it is easy to see that $H$ is the Noether charge for this time evolution. In gravity and field theory we will run this backwards: we trade $\dot x$ for an infinitesimal gauge transformation and calculate the right-hand side, which serves to \emph{define} the variation of the corresponding conserved charge $\delta H$.

\subsection{Noether charges of double field theory}

The covariant phase space approach applies to any theory formulated in terms of a variational principle. We will now apply it to double field theory. This leads to the conserved charges studied in \cite{Blair:2015eba , Park:2015bza ,Naseer:2015fba}. 

The variation of the DFT action (in this subsection we drop the $1/16\pi G_{DFT}$ prefactor to simplify expressions) reads
\be
\begin{split}
\delta S_{DFT} & = 
\int e^{-2d} \left( - 2 \delta d \mathcal{R} + \delta H^{MN} \mathcal{R}_{MN} \right) \\
 & + \int \partial_M \left( e^{-2d} \Theta^M \left [ H, d ; \delta H , \delta d \right] \right) \,,
\end{split} 
\ee
where the bulk term gives the equations of motion:
\be
\mathcal{R} = 0 
\quad,\quad
\mathcal{R}_{MN} = 0\,.
\ee
The total derivative term defines the \emph{symplectic potential}
\begin{align}
\Theta^M=\delta H^{QP} \left(\frac{1}{4} H^{MR}\partial_R H_{QP}-\frac{1}{2}H^{MR} \partial_Q H_{RP} - \frac{1}{2} H^R_Q \partial_R H^M_P\right) + 2 \delta H^{MP} \partial_P d\nonumber \\
-\partial_P\delta H^{MP}    +4\partial_P (\delta d)\:  H^{MP}\,,
\label{sympot} 
\end{align}
which can be explicitly checked to be a generalised vector under generalised diffeomorphisms.  

We will use the symplectic potential to define the symplectic form in a moment. Before we do that, we consider the variation of the DFT action under a generalised diffeomorphism with parameter $\Lambda^M$. This is a gauge invariance, so we only get a boundary term:
\be
\delta_\Lambda S_{DFT} = \int \partial_M \left( \Lambda^M e^{-2d} \mathcal{R} \right)\,.
\ee
By comparing the two variations it follows that the following current
\be
\label{DFT_current}
\mathcal{J}^M = e^{-2d} \left( \Theta^M [ H,d;\mathcal{L}_\Lambda H, \mathcal{L}_\Lambda d ] - \Lambda^M \mathcal{R} \right)
\ee
is divergence-free whenever $(d,H)$ are on-shell:
\begin{align}
\mathcal{R} = 0 = \mathcal{R}_{MN} \Rightarrow \partial_M\mathcal{J}^M=0\,.
\end{align}
Therefore on-shell there exists \cite{wald1990identically}, possibly only locally, an antisymmetric $\mathcal{J}^{MN}$ that satisfies
\begin{align}
\label{onshell_current_prepotential}
\mathcal{J}^M= \partial_N\mathcal{J}^{MN}\,.
\end{align}
We will see that $\mathcal{J}^{MN}$ integrated against a codimension 2 surface at infinity contributes to the Noether charge associated with $\Lambda^M$.

Let us write down Hamilton's equation in the covariant phase space form for the dynamics generated by the generalised diffeomorphism with parameter $\Lambda^M$:
\be
\slashed{\delta} Q_\Lambda=\Omega[d, H;(\delta d,\delta H), (\mathcal{L}_\Lambda d, \mathcal{L}_\Lambda H)]\,.
\ee
We will view this as a \emph{definition} of the infinitesimal Noether charge $\slashed{\delta} Q_\Lambda$ associated to $\Lambda^M$; we use a slashed delta notation because the existence of a $Q_\Lambda$ whose variation equals the right-hand side is in fact not guaranteed; we will elaborate on this later in this section. The symplectic form on the right-hand side is the integral of
\be
\Omega^M \left[ (d,H) ; (\delta_1 d , \delta_1 H) , (\delta_2 d , \delta_2 H) \right]
 = 
\delta_1 \left[ e^{-2d} \Theta^M( H,d;\delta_2 H , \delta_2 d ) \right]
-  \delta_2 \left[ e^{-2d} \Theta^M( H,d;\delta_1 H , \delta_1 d ) \right]\,.
\ee
If we specialise to the case where $\delta_2$ is an infinitesimal generalised diffeomorphism, it is not difficult to calculate
\be
\label{variational_identity}
\Omega^M \left[ (d,H) ; (\delta d , \delta H) , (\mathcal{L}_\Lambda d , \mathcal{L}_\Lambda H) \right]
 = \delta \mathcal{J}^M - \partial_P \left( e^{-2 d} (2 \Theta^{[M} \Lambda^{P]})\right) -e^{-2 d}\partial^M \Lambda_P \Theta^P
 \ee
where
\be
\Theta^M=\Theta^M \left [ H, d ; \delta H , \delta d \right]
\ee
assuming that
\begin{itemize}
\item the background fields $(d,H)$ are on-shell,
\item the generalised diffeomorphism parameter $\Lambda^M$ does not depend on the background $(d,H)$ and
\item $\Theta^M$ transforms as a generalised vector.
\end{itemize}
Note that the final term in \eqref{variational_identity} is of the form $(\dots)\partial^M ( \dots )$ and so is ``derivative-index valued'' in the language of \cite{Park:2015bza}. As a result, when such a term is integrated over a generalised hypersurface, as we explain in appendix \ref{appendix_integration}, it drops out by the section condition. In what follows, we will frequently drop such terms from e.g. the expressions for the current. 

The existence of a Noether charge $Q_\Lambda$ whose variation equals the right-hand side of \eqref{variational_identity} is equivalent to the existence of the ``boundary vector''\footnote{This is not necessarily a generalised vector. This non-covariance of the boundary term comes up generally in the covariant phase space approach, see the discussion in \cite{Iyer:1994ys} in section 6 around formulas (80)--(99). Physically speaking, one expects certain diffeomorphisms to change the values of the Noether charge integrals: consider e.g. the time-translation charge of a Schwarzschild black hole before and after a diffeomorphism realising a Lorentz boost in standard coordinates.} $B^M$
such that
\be
\label{boundary_vector_condition}
\delta\int_{\partial C(\infty)}e^{-2 d} (2 B^{M} \Lambda^{P}) \varepsilon_{MP}=\int_{\partial C(\infty)} e^{-2 d} (2 \Theta^{[M} \Lambda^{P]})\varepsilon_{MP}=\int_C\partial_P \left( 2 e^{-2 d} \Theta^{[M} \Lambda^{P]}\right) \varepsilon_M\,,
\ee
where the $\varepsilon$ are the normal and binormal to the codimension 1 ``Cauchy surface'' $C$ and its boundary at infinity $\partial C(\infty)$ respectively and we have used a Stokes' theorem for doubled spacetime in the third line; see Appendix \ref{appendix_integration}. When $B^M$ does exist the Noether charge is
\begin{align}
Q_\Lambda&=\int_C \mathcal{J}^M \varepsilon_M-\int_{\partial C(\infty)}e^{-2 d} (2 B^{M} \Lambda^{P}) \varepsilon_{MP}\\
&=\int_{\partial C(\infty)} \left( \mathcal{J}^{MP} -2e^{-2 d}B^{M} \Lambda^{P}\right)\varepsilon_{MP}\,.\label{noetherchargedefinition}
\end{align}
To get the final line we have used the fact $\mathcal{J}^M$ is divergence-free on-shell as well as Stokes' theorem.

Thus we see that the following antisymmetric generalised tensor density
\be
\label{charge_integrand}
\mathcal{Q}^{MP}=\mathcal{J}^{MP} -2e^{-2 d}B^{[M} \Lambda^{P]}\,,
\ee
integrates to define a conserved charge in DFT. The expression for $\mathcal{J}^{MN}$ has been determined to be \cite{Blair:2015eba, Park:2015bza}
\be
\label{current_formula}
e^{2d} \mathcal{J}^{MN}=2 (\partial_P \Lambda^{[M} + \partial^{[M}\Lambda_P) H^{N]P} +\Lambda^P(2\partial^{[M} H^{N]}_P + 2 H_{RP} H^{Q[M} \partial_Q H^{N] R} -H^Q_P H^{[M}_R \partial_Q H^{N]R})
\ee
using $\eta_{MN}$ to raise/lower indices.

The boundary vector can be taken to be as in \cite{Berman:2011kg, Park:2015bza, Naseer:2015fba}:
\be
\label{boundary_vector_def}
B^M \equiv  -\partial_P H^{M P}+ 4H^{MP} \partial_P d
\ee
which varies into $\Theta^M$ on the boundary where Dirichlet boundary conditions $\delta d=\delta H_{MN}=0$ hold.

\subsection{The first law of black hole thermodynamics}

In the Lee-Iyer-Wald covariant phase space formalism the first law of black hole thermodynamics is derived from a variational identity which sets the infinitesimal Noether charges of the previous section (which are integrals at spatial infinity) equal to an integral over the horizon; the last integral is proportional to the variation of the entropy, plus any charge contributions if the solution is supported by non-vanishing gauge fields.

To derive this identity, let us return to \eqref{variational_identity}. So far we have not imposed any conditions on the two variations $(\delta d, \delta H)$ and $(\mathcal{L}_\Lambda d, \mathcal{L}_\Lambda H)$. If we restrict to $(\delta d, \delta H)$ that solve the linearised equations of motion, it follows that there exists an antisymmetric $\delta \mathcal{J}^{MN}$ so that\footnote{Proof: Consider \eqref{onshell_current_prepotential} for some one-parameter family of \emph{on-shell} backgrounds, then take the variation.}
\be
\delta \mathcal{J}^M=\partial_N \delta\mathcal{J}^{MN}\,.
\ee
If in addition we consider a gauge parameter $\Lambda^M$ which is \emph{generalised Killing}, the left-hand side of \eqref{variational_identity} vanishes as it is linear in $(\mathcal{L}_\Lambda d, \mathcal{L}_\Lambda H)$ and after using the definition \eqref{charge_integrand} we obtain
\be
\label{1stlawdifferential}
\boxed{0=\partial_P \delta\mathcal{Q}^{MP} \,.}
\ee
Once we specify $\Lambda^M$ appropriately, equation \eqref{1stlawdifferential} is the first law of black hole thermodynamics in a ``differential'' form, stating that $\delta\mathcal{Q}^{MP}$ is conserved.

The standard form of the first law relates variations of the entropy to those of the mass, angular momentum, electric charge and other physical charges. In double field theory, as everything has been subsumed into the generalised metric (and dilaton, which does not play much of a role here), there is just the single Noether charge given by integrating $\mathcal{Q}^{MN}$. 

Let us assume that we have a background for which there exists a horizon specified by $R=R_0$ for a radial coordinate $R$. Then if we integrate \eqref{1stlawdifferential} against the codimension 1 ``Cauchy surface'' $C$ given by $t=t_0$  we obtain using Stokes' theorem 
\be
\begin{split} 
\int_{t=t_0,R =R_0}   d^{2(D - 2)} X & \: (\delta \mathcal{J}^{MN} - e^{-2 d} 2\Theta^{M}(\delta)\Lambda^{N})\varepsilon_{MN} \\
&   = \int_{t=t_0,R =+\infty}  d^{2(D-2)} X\: (\delta \mathcal{J}^{MN} - e^{-2 d} 2\Theta^{M}(\delta)\Lambda^{N})\varepsilon_{MN}  \,.
\end{split} 
\label{1stlawintegral1notsplit}
\ee
where as before $\varepsilon_{MN}$ is the binormal to the codimension 2 surface defined by $t=t_0$ and $R=R_0$ or $R=+\infty$ respectively, see appendix \ref{appendix_integration}.

To define precisely what we mean by ``horizon'' as well as identify distinct entropy, mass, winding charge etc. contributions in \eqref{1stlawintegral1notsplit}, we need a way to partially break $O(D,D)$.
This is provided by the split parametrisation of DFT introduced in \cite{Hohm:2013nja}. This rewrites DFT in terms of the variables which naturally appear in a Kaluza-Klein reduction, but without actually carrying out the full truncation. In this parametrisation, one has access to a conventional spacetime metric (in the non-dualisable, ``external'' dimensions), with respect to which one can define standard spacetime structures, such as a Killing horizon; the horizon thus defined could then be said to lie purely within the external space, although --- as we will see in section \ref{section_1stlaw_boundaryconditions} where the configurations we consider are characterised --- this language is somewhat misleading insofar as it implies the spacetimes under consideration are direct products of the external and internal spaces. The virtue of this definition is that if a solution has a horizon, then so do all its duals, since dualities only act on the doubled, internal geometry and not on the external geometry.

The other motivation for considering this parametrisation of double field theory is that it rewrites the theory in the same form as exceptional field theory \cite{Hohm:2013pua}, the U-duality invariant generalisation of double field theory. Thus, this parametrisation will teach us what to expect when we come to generalise our results from T- to U-duality.

\section{Split parametrisation and the first law} 
\label{split} 

\subsection{Decomposition of DFT and the current} 
\label{subection_splitdecomposition}
The split parametrisation that we will use is that introduced in \cite{Hohm:2013nja}. In this subsection, we first explain this parametrisation, and then give the expressions for how the components of the DFT Noether current decompose. 

The idea is to start with the usual $O(D,D)$ DFT, with coordinates $X^{\hat{M}}$, generalised metric $\hat{H}_{\hat M \hat N}$ and dilaton $\hat{d}$ (here we have introduced hatted $2D$ indices and fields in order to make the decomposition clearer). Then, one groups the coordinates into ``external'' and ``internal'' sets. The external coordinates and their duals are written $(x^\mu, \tilde x_\mu)$, with $\mu = 1,\dots d$, while the internal coordinates are written as $X^M$, with $M$ being now a fundamental $O(n,n)$ index with $n \equiv  D-d$. We impose the partial section condition solution, $\tilde \partial^\mu = 0$, so that the duals to the external coordinates do not appear, and maintain $\partial_M \cdot \partial^M \cdot = 0$ as the (formally unsolved) section condition on the internal doubled coordinates. Thus, altogether we have $X^{\hat M} = ( x^\mu, \tilde x_\mu, X^M)$, with the $\tilde x_\mu$ never appearing. 

The fields and gauge symmetries decompose in the same manner. This is similar to what one does in a Kaluza-Klein split, except that (aside from truncating the dependence on $\tilde x_\mu$) we do not perform a reduction. This is also entirely analogous to the manipulations carried out on supergravity when establishing the relationship to exceptional field theory \cite{Hohm:2013pua}. 

The generalised $\hat{H}_{\hat M \hat N}$ decomposes to produce an external metric and $B$-field, $g_{\mu\nu}$ and $B_{\mu\nu}$, an external one-form $A_\mu{}^M$ transforming in the fundamental of $O(n,n)$, and an internal generalised metric $H_{MN}$ parametrising the coset $O(n,n) / O(n) \times O(n)$ (we assume that time lies in the external directions). The explicit decomposition of the components is: 
\cite{Hohm:2013nja}
\begin{align}
\label{KK_parametrisation} 
\hat H_{\mu \nu} & = g_{\mu \nu} + g^{ \rho \sigma} C_{\mu  \rho} C_{\nu \sigma} + H_{M N} A_\mu{}^M A_\nu{}^N \,, & 
\hat H_{\mu}{}^\nu & = - g^{\nu  \rho} C_{\mu  \rho} \,,\\
\hat H^{\mu \nu} & = g^{\mu \nu} \,, & 
\hat H^{\mu}{}_M & = -g^{\mu  \rho} A_{ \rho M}\,, \\
\hat H_{\mu M} & = H_{MP} A_\mu{}^P + C_{\mu  \rho} g^{ \rho \sigma} A_{\sigma M}\,, & 
\hat H_{MN} & = H_{MN} + g^{ \rho \sigma} A_{ \rho M} A_{\sigma N} \,,
\end{align} 
where the doubled internal index on $A_\mu{}^M$ is now lowered with the $O(n,n)$ structure $\eta_{MN}$, and
\be
C_{\mu \nu} \equiv - B_{\mu \nu} +\frac{1}{2}  A_\mu{}^M A_{\nu M}\,.
\ee
The $O(D,D)$ generalised dilaton is rewritten as
\be
e^{-2\hat d}= \sqrt{\det g} e^{-2d}\,.
\ee
where $e^{-2d}$ is now the $O(n,n)$-invariant generalised dilaton. If we were to truncate all dependence on the internal doubled coordinates, then we would arrive at the Kaluza-Klein reduction of the NSNS action to $d$ dimensions, with $g_{\mu\nu}$  a string frame metric. We also mention our conventions involve using minus the $B$-field of \cite{Hohm:2013nja}.

We observe that this parametrisation interpolates between the fully $O(D,D)$ covariant DFT formalism (for the number of external dimensions $d=0$) and the standard formulation of 
low energy string theory \eqref{spacetime_action} (for $d=D$).

The $O(D,D)$ generalised diffeomorphism generator $\hat \Lambda ^{\hat M}$ splits into an external diffeomorphism $\xi^\mu$, an external $B$-field gauge transformation $\lambda_\mu$, and an $O(n,n)$ generalised diffeomorphism $\Lambda^M$ as
\be
\Lambda^{\hat M}=(\xi^\mu, \lambda_\mu, \Lambda^M)\,.
\ee
The transformations that result take a somewhat intricate structure, and may be perused in appendix \ref{splittransfs}.

The gauge fields $A_\mu{}^M$ and $B_{\mu\nu}$ constitute the ``tensor hierarchy'' of the split theory \cite{Hohm:2013nja,Wang:2015hca}. Their field strengths are
\be
\begin{split} 
\mathcal{F}_{\mu \nu}{}^M & = \partial_\mu A_\nu{}^M - \partial_\nu A_\mu{}^M - [ A_\mu, A_\nu ]_C^M + \partial^M B_{\mu \nu}\,,  \\
\end{split} 
\ee
\be
\begin{split} 
\mathcal{H}_{\mu \nu \rho} & = 
 3 D_{[\mu } B_{\nu \rho]}
-3 A_{[\mu}{}^N \partial_\nu A_{\rho]N}  
+ A_{[\mu |N|} [ A_\nu, A_{\rho]} ]_C^N \,, 
\end{split} 
\ee
with the internal $C$-bracket defined as in \eqref{courant}. These field strengths are invariant under the gauge transformations \eqref{eq:deltalambda} and transform as generalised tensors under generalised diffeomorphisms, consistent with their index structure. 

In the above, we have introduced the  derivative $D_\mu = \partial_\mu - \mathcal{L}_{A_\mu}$, which is covariant under generalised diffeomorphisms, as explained in \cite{Hohm:2013nja}. One has e.g. $D_\mu g_{\nu \rho} = \partial_\mu g_{\nu \rho} - A_\mu{}^N \partial_N g_{\nu \rho}$ as $g_{\mu \nu}$ is a scalar under generalised diffeomorphisms. Note that one also has $D_\mu B_{\nu \rho} = \partial_\mu B_{\nu \rho} - A_\mu{}^N \partial_N B_{\nu \rho}$.

The above results summarise the essential features of the split parametrisation that we require. We can now work out the form of the current in this version of the theory: inserting the above decompositions \eqref{KK_parametrisation} into \eqref{current_formula}, a laborious calculation\footnote{The computer algebra program Cadabra \cite{Cadabra, Peeters:2007wn} proved useful here.} gives the components of the current. 
The purely external components are all that we will actually need, and are given by
\be
\begin{split}
e^{2\hat d} J^{\mu \nu} & = 
2 \underline{\nabla}^{[\nu} \xi^{\mu]} 
- \tilde \lambda_\rho \mathcal{H}^{\mu \nu \rho} 
- \tilde \Lambda^M H_{MN} \mathcal{F}^{\mu \nu N }\,,
\end{split} 
\label{Jext}
\ee
where
\begin{align}
\label{thermo_potential_def}
\tilde \lambda_\mu  \equiv  \lambda_\mu - \xi^\lambda C_{\lambda \mu} - \Lambda^P A_{\mu P}\,, \qquad \tilde \Lambda^M  \equiv  \Lambda^M + \xi^\lambda A_{\lambda}{}^M
\end{align}
and the Levi-Civita connection $\underline{\nabla}_\mu$ is built using $D_\mu$ and $g_{\mu \nu}$ with Christoffel symbol
\be
\underline{\Gamma}_{\rho\mu}^\nu \equiv  \frac{1}{2} g^{\nu\sigma} ( D_\mu g_{\sigma \rho} +D_\rho g_{\sigma\mu} - D_\sigma g_{\rho\mu})\,.
\ee
Also, the boundary vector \eqref{boundary_vector_def} components are
\be
\label{bvext}
B^\mu = -D_\nu g^{\mu\nu} - g^{\mu\nu} D_\nu\ln g
+4 g^{\mu\nu} D_\nu d 
+g^{\mu\nu}  \partial_N A_\nu{}^N \,.
\ee
For completeness, we also record the other components of \eqref{current_formula} and the boundary vector in the appendix \ref{Japp}.

\subsection{The first law and duality-invariant entropy and mass formulas} 
\label{section_1stlaw_boundaryconditions}

We now consider the first law in this decomposition. We shall see that using the split form of DFT gives us more control over the definition of the horizon, and leads quite naturally to a T-duality invariant definition of black hole entropy.

We begin by making some assumptions on the form of the backgrounds we will consider. We assume that the $d$ external coordinates are $x^\mu=(t, x^i)$, $i=1,\dots,(d-1)$, so that in particular they include time (the $x^i$ will be interpreted shortly as (asymptotically) Cartesian coordinates). We then impose conditions on the fields, including most importantly an asymptotic flatness condition on $g_{\mu\nu}$ as the radial coordinate $R \equiv \sqrt{x^i x^j \delta_{ij}} \rightarrow \infty$. These conditions are:
\begin{itemize}
\item The external metric $g_{\mu\nu}$ is static (for simplicity; we sketch the generalisation to stationary $g_{\mu\nu}$ at the end of the subsection) and asymptotically flat as $R\rightarrow \infty$ in the asymptotically Cartesian coordinate system $(t,x^i)$. We assume that $g_{\mu\nu}$ has a normalised asymptotically timelike Killing vector $\xi^\mu = \partial/\partial t$, so that $\xi^2  \equiv  \xi^\mu\xi^\nu g_{\mu\nu} \rightarrow -1$ for $R \rightarrow \infty$, and further that there is a bifurcate Killing horizon (see e.g. \cite{Racz:1992bp} for a definition) for $\xi^\mu$ at $R=R_0$, with constant non-zero surface gravity $\kappa$ (defined below). We will also assume that $g_{\mu\nu}$ is independent of the internal doubled coordinates.
\item The gauge fields $A_\mu{}^M$ and $B_{\mu\nu}$ vanish for $R \rightarrow \infty$. 
\item The generalised metric $H_{MN}$ goes to the $2n\times 2n$ identity matrix $\delta_{MN}$, with the generalised dilaton similarly going to 1.
\item Finally, we require that $\xi^\mu$ is generalised Killing acting on the above fields in addition to the metric (i.e. the right-hand sides of \eqref{eq:extdiffeos} all vanish). This is trivially satisfied if $\xi^\mu = \partial/\partial t$ and all fields are $t$-independent.
\end{itemize} 
From a $D$-dimensional perspective these assumptions can accommodate both asymptotically flat $(\mathbb{R}^{1,D-1})$ and product geometries $(\mathbb{R}^{1,d-1}\times T^n)$ depending on whether the $n$ internal doubled coordinates are assumed to be compact, as $H_{MN}=\delta_{MN}$ describes either. Our assumptions on $g_{\mu\nu}$ in particular further imply that the surface gravity $\kappa$ which we define by 
\begin{align}
\partial_\mu (\xi^\sigma \xi^\nu g_{\sigma\nu}) = (-2 \kappa) \xi_\mu &&\text{on the Killing horizon }
\end{align}
is constant along internal directions of the horizon and the bifurcation surface and is invariant under $O(n,n)$ dualities, and the results of Racz and Wald \cite{Racz:1992bp} suggest that it is sensible to consider only the case where $\kappa$ is constant along external directions as well. We finally quote the following standard result (see e.g. \cite{Mohaupt:2000mj}) for later use
\be
\label{nabla_xi_identity}
\nabla_{\mu} \xi_\nu = \kappa \epsilon_{\mu\nu}\,,\qquad \epsilon_{\mu\nu} \epsilon^{\mu\nu}=-2\,,
\ee
valid on the bifurcation surface, where $\epsilon_{\mu\nu}$ is proportional to, but must not be confused with, the binormal $\varepsilon_{\mu\nu}$ defined in appendix \ref{appendix_integration} (n.b. the different normalisation).

We stress that the assumptions in the previous paragraph \emph{do not} entail that the field configurations under consideration are direct products of non-dualisable external and doubled internal spaces: while the external metric $g_{\mu\nu}$ indeed does not depend on the internal coordinates, other fields depend on all coordinates (excepting, of course, coordinates corresponding to (generalised) Killing vectors); for instance the internal generalised metric $H_{MN}$ is allowed to depend on external coordinates, and indeed it must if it is to satisfy the boundary condition. We will see an example in the discussion of the black string solution later.

With the above definition and setup, it is trivial to observe that the $O(n.n)$ dual of any solution possessing a horizon also has a horizon, since the $O(n,n)$ subgroup does not act on the external metric or the external coordinates. We can now also clarify the nature of the ``Cauchy surface'' $C$ we have been using: $C$ is simply the level set $t=t_0$ where $t$ is the timelike external coordinate (one could of course consider the level set of any scalar on external space). This defines a $(d-1)$-dimensional hypersurface on the external space. By making a choice of ($n$-dimensional) section for the doubled internal space, we get a $(d-1)+n=(D-1)$-dimensional submanifold inside the physical, undoubled spacetime.

We now return to the first law, which was written previously in the form \eqref{1stlawintegral1notsplit}. 
Since the 
$t,R$ coordinates are both external, $\varepsilon_{\hat M \hat N}$ is only non-vanishing in its external components and we have $\delta \mathcal{Q}^{\hat M \hat N}\varepsilon_{\hat M \hat N}=\delta \mathcal{Q}^{\mu\nu}\varepsilon_{\mu\nu}$.

We evaluate the expressions in terms of the \emph{unit surface gravity} Killing vector 
\be
\xi' \equiv  \frac{1}{\kappa} \xi
\ee
where $\xi = \partial/\partial t$.\footnote{\label{unitsurfacegravityfootnote}We do this because it is consistent to set $\delta \xi'=0$ but not to set $\delta \xi=0$ (i.e. the $\xi^\mu$ we use in the main text is field-dependent), as mentioned in Wald's original \cite{Wald:1993nt} and explained in detail in \cite{Mukohyama:1998mq}; as the subtlety is not intrinsic to double field theory we will not elaborate here. Since, as we find below, the Noether charge for $\xi'$ at the horizon is just twice the black hole area, using $\xi'$ gives the first law in the form \be 2 \delta A= \delta_{\infty} M/\kappa + \dots \ee which is equivalent to the usual form as long as $\kappa \neq 0$. The modified variation $\delta_\infty$ is defined in Mukohyama's work \cite{Mukohyama:1998mq} and accounts for modifications to the Killing vectors arising from variations of the geometry.} 
We then find 
\be
\begin{split} 
\frac{1}{16\pi G_{DFT} } \int_{t=t_0,R =R_0}   d^{d-2}x d^{2n}X & \: (\delta \mathcal{J}^{\mu\nu} - e^{-2 \hat d} 2\Theta^{\mu}(\delta)\xi'^{\nu})\varepsilon_{\mu\nu} \\
&   = \frac{1}{16\pi G_{DFT}}  \int_{t=t_0,R =+\infty} d^{d-2}x d^{2n}X\: (\delta \mathcal{J}^{\mu\nu} - e^{-2 \hat d} 2\Theta^{\mu}(\delta)\xi'^{\nu})\varepsilon_{\mu\nu}  \,,
\end{split} 
\label{1stlawintegral1}
\ee
where $\mathcal{J}^{\mu\nu}$ was given in \eqref{Jext}, and $\Theta^\mu(\delta)$ is a component of the symplectic potential \eqref{sympot}. We have reinstated the $1/16\pi G_{DFT}$ prefactor: this is defined as in \eqref{eq:GDFT}, though we now only have duals for $n$ directions. 

Let us first consider the terms at infinity. The staticity assumption and falloff conditions on the gauge fields imply there is only one term, corresponding to the variation of the Noether charge associated to translation in (asymptotic) time. Time translation is generated by the normalised asymptotically timelike Killing vector $\xi$ (with $\xi^2\to -1$), so this term is identified as the variation of the energy, which equals the mass variation since the black hole is not rotating in the external dimensions. We thus define mass as the Noether charge \eqref{noetherchargedefinition}:
\be
\label{massdef}
 M  \equiv  Q_\xi = \frac{1}{16 \pi G_{DFT} } \int_{t=t_0,R = + \infty}  d^{d-2}x d^{2n}X\: (\mathcal{J}^{\mu\nu} - e^{-2 \hat d} 2B^{\mu}\xi^{\nu})\varepsilon_{\mu\nu}  \,.
\ee
This enters the first law through the variation of the Noether charge for the unit surface gravity Killing vector $\xi'$. Taking into account the remark in footnote \eqref{unitsurfacegravityfootnote}, the right-hand side of \eqref{1stlawintegral1} is
\be
\frac{1}{\kappa}\delta M \equiv  \delta Q_{\xi'}\,.
\ee
Note that the mass definition is the only one where the boundary vector \eqref{boundary_vector_def} makes a contribution (through its external component \eqref{bvext}).

Now we turn to the horizon contributions. Here we will get a linear combination of variations of the entropy (the Noether charge of $\xi^\mu$ there) and of electric charges associated to the gauge fields. Stationarity of the background implies we can evaluate the left-hand side of \eqref{1stlawintegral1} on any horizon cross-section \cite{Jacobson:1993vj,Iyer:1994ys}, and it is convenient to do so on the bifurcation surface\footnote{However, note that the bifurcation surface always lies outside of the $(t,x^i)$ external coordinate chart we are using. With this caveat understood there is no need to introduce a new chart, valid where $\xi^\mu=0$.}, where the Killing vector $\Lambda'^{\hat M}=(\xi')^\mu$ vanishes. Since $\Theta(\delta)$ is linear in variations it is finite everywhere and the second term in the left-hand side of \eqref{1stlawintegral1} vanishes, leaving us with
\be
\frac{1}{16\pi G_{DFT}} 
\int_{t=t_0,R =R_0}  d^{d-2}x d^{2n}X\: \delta \mathcal{J}^{\mu\nu}\varepsilon_{\mu\nu}\,.
\ee
Therefore we are looking at the variation of
\be
\begin{split}
 \mathcal{J}^{\mu \nu} & = e^{-2\hat d}\left( -2 g^{  \rho[\mu}\nabla_ \rho (\xi')^{\nu]}  
- \tilde \lambda_  \rho \mathcal{H}^{\mu \nu   \rho} 
- \tilde \Lambda^M H_{MN} \mathcal{F}^{\mu \nu N } \right)\varepsilon_{\mu\nu}\,.
\end{split}
\ee
on the bifurcation surface, where
\be
\tilde \lambda_\mu \equiv  - (\xi')^\rho C_{ \rho \mu}\,, \qquad \tilde\Lambda^M \equiv  (\xi')^ \rho A_\rho{}^M \,.
\ee 
There are three terms:
\begin{itemize}
\item The ``Komar'' term $\delta \left( -2e^{-2\hat d} g^{  \rho[\mu}\nabla_  \rho (\xi')^{ \nu]}\right)\varepsilon_{\mu\nu}$ contributes the entropy variation. Consider the integral
\be
Q_{hor, \xi'}  \equiv   \frac{-2}{16\pi G_{DFT}}  \int_{t=t_0,R=R_0}  d^{d-2}x d^{2n}X\: e^{-2\hat d} g^{  \rho[\mu}\nabla_ \rho (\xi')^{\nu]}\varepsilon_{\mu\nu}\,.
\ee 
Using \eqref{nabla_xi_identity} (replacing $\kappa$ by $1$ for the unit surface gravity $\xi'$) and \eqref{determinantmagic} for the external metric the integrand is rewritten as
\be
-2 e^{-2\hat d} g^{  \rho[\mu}\nabla_ \rho (\xi')^{\nu]}\varepsilon_{\mu\nu}=-2e^{-2d} \sqrt{|\det g|} \epsilon^{\mu\nu}\varepsilon_{\mu\nu}=2 e^{-2d} \sqrt{|\det g_{d-2}|}
\ee
where $\sqrt{|\det g_{d-2}|}$ is the determinant of the $(d-2)\times(d-2)$ external metric induced on the bifurcation surface $(t=t_0, R=R_0)$\footnote{Recall that $t,R$ were both designated external and also that $\epsilon$ is defined in \eqref{nabla_xi_identity} whereas $\varepsilon$ is defined in appendix \ref{appendix_integration} and is such that $\varepsilon_{\mu\nu} = \delta^t_{[\mu} \delta^R_{\nu]}$ in our coordinates.}). Thus
\be
Q_{hor, \xi'}= \frac{1}{8\pi G_{DFT}} \int_{t=t_0,R=R_0}  d^{d-2}x d^{2n}X\: e^{-2d} \sqrt{|\det g_{d-2}|}\,.
\ee
If one expresses the fields of the split parametrisation in terms of the usual spacetime fields using formulas \eqref{split_to_spacetime}, and also solves the strong constraint in the usual way ($\tilde\partial^{\hat \mu}=0$), one finds
\be
Q_{hor,\xi'} = \frac{1}{8 \pi G} A \,,
\ee 
where $A$ is the horizon area in the Einstein-frame spacetime ($D$-dimensional) metric, and $G$ is Newton's constant in spacetime. The familiar $S = A/4G$ Bekenstein-Hawking entropy formula suggests we are therefore more generally led to identify the entropy $S$ with
\be
\label{entropy_def}
S  \equiv 2 \pi Q_{hor, \xi'} =  \frac{1}{4 G_{DFT} } \int_{t=t_0,R=R_0}  d^{d-2}x d^{2n}X\: e^{-2d} \sqrt{|\det g_{d-2}|}\,.
\ee
This expression is manifestly $O(n,n)$-invariant and agrees with that derived from \eqref{spacetime_action}; it is, however, strictly more general since it is valid for \emph{any} parametrisation of the internal fields and choice of internal section. The physical interpretation in this more general scenario is supported by the appearance of $S$ in the first law of black hole thermodynamics, which we will show shortly; for now we write
\be
\delta Q_{hor,\xi'} = \frac{1}{2\pi} \delta S \,,
\ee
which is little more than a definition of some quantity $S$. 

\item The other two terms contribute what would be a mix of $B$-field charge and momenta from a $D$-dimensional spacetime perspective. Since $\xi'=0$ on the bifurcation surface and $\delta \xi'=0$ everywhere, and the variations $\delta C_{\mu \nu},\delta A_\mu{}^M$ do not diverge\footnote{The rationale being that if they did diverge, they would not be infinitesimal. The gauge fields themselves, however, generally \emph{do} diverge on the bifurcation surface so that $\tilde \lambda_\mu$ and $\tilde\Lambda^M$ are finite there, see \cite{Copsey:2005se,Keir:2013jga}.}, formula \eqref{thermo_potential_def} implies
\be
\delta \tilde \lambda_\mu=\delta \tilde\Lambda^M=0 \qquad \text{(on the bifurcation surface)}\,.
\ee
This suggests $\tilde \lambda_\mu$ and $\tilde\Lambda^M$ should be closely related to the thermodynamically conjugate variables multiplying the variations of the corresponding electric charges in the first law. This can be shown to be true for ordinary p-form gauge fields on spacetime using Poincar\'e duality (see \cite{Keir:2013jga}). However this is not available in the current setting so we will simply write the contribution of these two terms as the $O(n,n)$ invariant thermodynamic work contribution:
\be
\label{eleccharges}
\slashed{\delta} W  \equiv  -\frac{1}{16\pi G_{DFT}} \int_{t=t_0,R=R_0}  d^{d-2}x d^{2n}X\: \left[\tilde \lambda_\rho\delta\left(e^{-2\hat d}\mathcal{H}^{\mu\nu  \rho}\right) +\tilde\Lambda^M \delta\left(e^{-2\hat d}H_{MN} \mathcal{F}^{\mu\nu N } \right)\right] \varepsilon_{\mu\nu}\,,
\ee
where $\tilde \lambda_\mu$ and $\tilde\Lambda^M$ on the right-hand side are now expressed \emph{in terms of the canonically normalised $\xi$ rather than $\xi'$} (this is now legal since $\xi'$ now appears outside the variations).
In appropriate coordinates $\tilde \lambda_\mu$ and $\tilde\Lambda^M$ can be ``pulled out'' of the integrals so as to exhibit the right-hand side as a linear combination of variations of electric charge integrals; we give an example of how this works in the next section.

\end{itemize}

Putting everything together, we obtain the first law of black hole thermodynamics in its usual form:
\be
\label{1stlawintegralfinal}
	\delta M=  \frac{ \kappa }{2 \pi}  \delta S + \slashed{\delta}W \,.
\ee
All terms are (individually and manifestly) $O(n,n)$-invariant. The variation $\delta$ is assumed to satisfy the linearised equation of motion, but is otherwise arbitrary; it can, in particular, be time-dependent. For $d=D$ it is easy to see that \eqref{1stlawintegralfinal} is equivalent to the first law of black hole thermodynamics for the standard (string-frame) metric, dilaton, and Kalb-Ramond fields of the type II theories: $\mathcal{F}^{\mu\nu M }$ vanishes identically, $\mathcal{H}_{\mu\nu\rho}$ reduces to the NS 3-form field strength, and \eqref{entropy_def} reduces to the area formula (with an extra dilaton factor because we are in the string frame) by a standard calculation (see e.g. \cite{Iyer:1994ys}). We need also the standard definition of the temperature as $T = \kappa / 2 \pi$. 

It is straightforward to generalise \eqref{1stlawintegralfinal} to the case where the external metric $g_{\mu\nu}$ is only stationary, rather than static, but this requires assuming --- or proving --- some sort of horizon rigidity theorem, valid in the current context of the split parametrisation of \cite{Hohm:2013nja}, that guarantees the existence of some number of commuting Killing vector fields $\partial/\partial \varphi^I$ so that $\xi=\partial/\partial t + \Omega^I \partial/\partial \varphi^I$; the left-hand side of \eqref{1stlawintegralfinal} would then be replaced by $\delta E-\Omega^I \delta J_I$ (where of course $E \equiv  Q_{\partial/\partial t}$ and $J_I \equiv  -Q_{\partial/\partial \varphi^I}$).

With that caveat understood, \eqref{1stlawintegralfinal} accounts for all diffeomorphism and electric charges. However, it does \emph{not} contain magnetic charge contributions. This is because we derived the first law through the conservation law \eqref{1stlawdifferential} of the \emph{Noether} charge associated to a generalised diffeomorphism $\Lambda^M$; magnetic charges, on the other hand, arise from topological conservation laws without associated gauge invariances, so they are not automatically taken into account using this method. This is an issue with covariant phase space methods in general. There does not appear to exist a straightforward way to remedy this at present, but we will provide some suggestions in the conclusions. 

\section{Example: the black string} 
\label{example} 

\subsection{The black string and T-duality} 

The black string solution in $D=10$ dimensions is \cite{Horowitz:1991cd} 
\be
\begin{split}
ds^2 & = H^{-1} \left( - W dt^2 + dz^2 \right) + W^{-1} dr^2 + R^2 d \Omega_7^2 \,, \\
B_{tz} & = \alpha \left( H^{-1} - 1\right) \,\\
e^{-2(\phi-\phi_0)} & = H \,,
\end{split}
\ee
where
\be
H = 1 + \frac{r_-^6}{R^6} \quad W = 1 - \frac{r_+^6 - r_-^6}{R^6} \quad \alpha = \frac{r_+^3}{r_-^3} \,.
\ee
The Killing vector 
\be
\xi = \frac{\partial}{\partial t} \,,
\ee
has a Killing horizon at $R^6 = r_+^6 - r_-^6$. 

One can carry out a Buscher duality on the $z$ direction. This gives another black string, now carrying momentum along the dual circle $\tilde z$: 
\be
\begin{split}
ds^2 & = - H^{-1} W dt^2 + H \left( d \tilde z + \alpha ( H^{-1} - 1) dt \right) ^2 +  W^{-1} dR^2 + R^2 d \Omega_7^2\,,\\
B & = 0 \,,\\
e^{-2(\phi-\phi_0)} & = 1 \,.
\end{split} 
\label{blackpp}
\ee
The Killing vector which becomes null on the horizon at $R^6 = r_+^6 - r_-^6$ is now
\be
\tilde \xi = \frac{\partial}{\partial t} + \frac{1}{\alpha} \frac{\partial}{\partial \tilde z} \,.
\ee
This is canonically normalised since the asymptotically timelike Killing vector field $\partial/\partial t$  on the right-hand side has norm $-1$ at infinity; we can thus identify $1/\alpha$ with the velocity of the string in the $\tilde z$ direction. In fact it is not hard to see that this solution is a Lorentz-boosted Schwarzschild$\times S^1$ \cite{Horne:1991cn}, where the rapidity $\psi$ is related to the parameter $\alpha$ by $\alpha=(\tanh\psi)^{-1}$.

One might wonder whether the string velocity appears in the dual solution. In fact, on the horizon, one has that $B_{tz}(R^6 =r_+^6 - r_-^6) = -1/\alpha$. One can view this as the electric potential for this field.  

Now, let us embed this pair of solutions into double field theory. The generalised metric can be specified by writing the formal expression $ds^2 = H_{MN} dX^M dX^N$ as follows:
\be
\begin{split}
ds^2 & = - H^{-1} W dt^2 + H \left( d \tilde z + \alpha ( H^{-1} - 1) dt \right) ^2 
\\ & \quad - H W^{-1} d \tilde t{}^2 + 2 \alpha(H^{-1}-1) HW^{-1} d\tilde t dz  + H^{-1} ( 1 - \alpha^2 ( H-1)^2 W^{-1} ) dz^2 
\\ & \quad +  W^{-1} dR^2 + R^2 d \Omega_7^2 + W d\tilde{R}^2 + R^{-2} d \tilde{\Omega}_7^2 \,.
\end{split} 
\label{genlinelt} 
\ee
In the extremal limit of $r_+ \rightarrow r_-$, one finds the double pp-wave of \cite{Berkeley:2014nza}.

From the generalised metric, using \eqref{KK_parametrisation}, one can read off the fields in the split form. Here, we take the internal doubled coordinates to be just the minimal pair of $z$ and $\tilde z$. Then, we have
\be
\begin{split}
ds^2 & = - H^{-1} W dt^2 + W^{-1} dR^2 + R^2 d\Omega_7^2 \,,\\
A_{t}{}^M & = \begin{pmatrix} 0 \\ \alpha(H^{-1} - 1 ) \end{pmatrix}  \,,\\
H_{MN} & = \begin{pmatrix} H^{-1} & 0 \\ 0 & H \end{pmatrix} \,, \\
 e^{-2d} & = H^{1/2}  \,,
\end{split} 
\ee
with no other non-zero fields. We see that the field strength of $A_\mu{}^M$ is 
\be
\mathcal{F}_{tR z} = - \alpha \partial_RH^{-1} \,.
\ee

\subsection{First law for the black string} 

We will now explicitly calculate the conserved charges and verify the first law \eqref{1stlawintegralfinal} for the black string solution. We will only consider variations of the black string metric parameters $r_\pm$. These induce stationary variations of the metric and other fields, so we have the freedom to evaluate all horizon integrals on \emph{any} horizon cross-section \cite{Jacobson:1993vj,Iyer:1994ys}, rather than just the bifurcation surface, which we will exploit without further comment.

We have set up our formalism so that we will be able to be agnostic about our choice of section. The doubled space has coordinates $X^M = (z,\tilde z)$. We will denote by $X$ the chosen physical coordinate, so that $X = z$ or $X = \tilde z$. We assume these parametrise dual circles, so that the radii will be either $R_X  \equiv  R_z$ or $R_X = R_{\tilde z} = \alpha^\prime / R_z$. We also write $G_X$ to denote the Newton's constant of the ($D=10$) supergravity action in the frame with coordinate $X$, and, applying the definition \eqref{eq:GDFT} for the case of a single doubled direction, we let $G_{DFT} = 2\pi R_{\tilde X} G_X = 2\pi R_X G_{\tilde X}$ , so that 
\be
\frac{1}{16 \pi G_{DFT}} \int d^2X = \frac{2\pi R_X}{16 \pi G_X} =  \frac{2\pi R_z }{16 \pi G_z}
= \frac{2\pi R_{\tilde z}}{16 \pi G_{\tilde z}} \,.
\ee
As both $R_X$ and $G_X$ change under T-duality, our expressions will be fully T-duality invariant.

Recall the DFT Noether charge is given by the integral of
$Q^{\mu\nu} = \mathcal{J}^{\mu\nu} + 2 e^{-2d} \sqrt{g} \xi^{[\mu} B^{\nu]}$ with $J^{\mu\nu}$ given by \eqref{Jext} and \eqref{thermo_potential_def} and the boundary vector, which only contributes at infinity, given by \eqref{bvext}.
We integrate this charge over a constant $t=t_0$ hypersurface and at either the horizon at $R = R_0 \equiv r_+^6 - r_-^6$, or at infinity. We have $\mathcal{Q}^{\mu\nu}\varepsilon_{\mu\nu}=\mathcal{Q}^{t R}$ and so we just need to consider the integrand $\mathcal{Q}^{tR}$. For the solution we are considering we have
\be
\mathcal{Q}^{tR} [ \xi, \Lambda ] 
= e^{-2d} \sqrt{|g|}  \big( g^{tt} g^{RR} \xi^t  \partial_R g_{tt} +\xi^t B^R\big)  
- e^{-2d} \sqrt{|g|} g^{tt} g^{RR} ( \Lambda_z + \xi^t A_{tz} ) H^{zz}\mathcal{F}_{tR z} \,,
\ee
with the arguments in the square brackets denoting which generalised Killing vectors are contributing.
The integrated charge, in turn, is denoted
\be
Q [ \xi, \Lambda ] = \frac{1}{16\pi G_{DFT}} \int d^7x d^2X \mathcal{Q}^{tR}[ \xi, \Lambda] \,.
\label{eq:thecharge}
\ee
We will specialise from now on to the case $\xi^t = 1$, corresponding to the timelike Killing vector $\xi^\mu = \partial/\partial t$, and $\Lambda_z = 1$, corresponding to the invariance under translations in the $\tilde z$ circle, generated by $\Lambda^M = \partial / \partial \tilde z$. Note that in the section where the solution carries $B$-charge, this corresponds to a generator of gauge transformations and is instead viewed as $\Lambda^M = dz$. The picture here is exactly the same as suggested in \cite{Berkeley:2014nza} and confirmed in \cite{Blair:2015eba, Park:2015bza, Naseer:2015fba}. 

Now, let us identify the charges carried by this solution. 

\subsubsection*{The electric charge} 

We define an ``electric'' charge
\be
q_{elec}  \equiv  Q [ 0 , \Lambda_z=1]
= \frac{1}{16\pi G_{DFT}} \int_{t=t_0,R=R_0}  d^{7}x d^{2}X \mathcal{Q}^{tR} [ 0 , \Lambda_z=1 ] \,,
\ee
where, for this solution, the integration could equally well be taken at infinity or any constant value of $R$. 
For the solution we are considering, we find
\be
q_{elec} = - \frac{1}{16\pi G_X}  6 \alpha r_-^6 (2 \pi R_X) \vol \,.
\ee
Recall that $\alpha = r_+^3 / r_-^3$.
In the original, ``charged'' frame, this is the actual $B$-field charge of the string. In the dual frame, this becomes the momentum around the dual circle. 

\subsubsection*{The entropy and horizon charges} 

The entropy is defined in equation \eqref{entropy_def}. We have
\be
\begin{split} 
S & = \frac{1}{4 \pi G_{DFT}} \int_{t=t_0,R=R_0} d^7 x d^2X  e^{-2d} \sqrt{ \det g_7 } \\
 & = \frac{1}{4 G_X} (2\pi R_X) \vol r_+^3 ( r_+^6 - r_-^6 )^{2/3} \,.
\end{split} 
\label{entropyF1}
\ee

This entropy enters into the full horizon charge associated to $\xi^t =1$ in the following manner. First, we can evaluate the surface gravity for the Killing vector $\xi = \partial/\partial t$, finding
\be
\kappa = \frac{3 ( r_+^6 - r_-^6 )^{1/3}}{r_+^3} \,.
\label{kappaF1}
\ee 
Then we can evaluate the charge \eqref{eq:thecharge} on the horizon where we have:
\be
\begin{split} 
16 \pi G_{DFT} Q [ \xi^t, 0 ]\Big|_{hor} & = 
\int d^7 x d^2X e^{-2d} \sqrt{g} g^{tt} g^{RR} \partial_R g_{tt} 
-  \int d^7 x d^2 X e^{-2d} \sqrt{g} g^{tt}g^{RR} A_{tz} H^{zz} \mathcal{F}_{tR z} \\
 & = 
6 (r_+^6 - r_-^6 ) (2 \pi R_X) \vol 
+ 6  r_-^6 (2 \pi R_X) \vol \,.
\end{split} 
\ee
Now, the value of $A_{tz}$ at the horizon is simply $-1/\alpha$. We identify this as an electric potential, $\Psi  \equiv  - 1/\alpha$ (in the original black string frame, $\Psi$ is indeed an electric potential difference for the $B$-field, but in the dual frame it equals the velocity of the horizon). Comparing this charge with the expressions \eqref{entropyF1} and \eqref{kappaF1}, one see that in fact
\be
Q [ \xi^t , 0 ]\Big|_{hor} 
= \frac{\kappa}{2\pi} S + \Psi q_{elec}   \,.
\ee

\subsubsection*{The mass} 

Finally, we want to determine the charge at infinity associated to $\xi = \partial/\partial t$ which gives the mass. 
Some care must be taken in evaluating the boundary vector contribution at infinity. As outlined in section \ref{section_1stlaw_boundaryconditions}, it is only defined for coordinate systems which are asymptotically Cartesian (because our definition \eqref{boundary_vector_def} requires the field variations to vanish at infinity). Suppose $x^\mu = ( t, x^i)$ is such a coordinate system, and let $r= |x_i x^i|$ be the norm squared of the Cartesian spatial coordinates used. 
Then (we again refer to appendix \ref{appendix_integration} for details on $\varepsilon$)
\be
\xi^\mu B^\nu \varepsilon_{\mu\nu}= \xi^\mu B^\nu \varepsilon_{\mu\nu}=\xi^\mu B^\nu\partial_{[\mu} t \partial_{\nu]} r= \xi^\mu B^\nu \partial_{[\mu} t\left(\frac{x^i}{r} \delta_{\nu]}^i \right)
\ee
which implies that $B^\mu$ only contributes through its ``radial'' component $B^r$ which we \emph{define} as $B^r  \equiv  B^i x^i/r$\,. Assuming that $g_{\mu\nu}$ is independent of the internal coordinates along with $d$ and $A_\mu{}^M$ --- as is the case here --- one obtains
\be
B^r \approx n^i \eta^{jk} \partial_j g_{ik} - \eta^{ij} \partial_r g_{ij}   + \partial_r g_{tt} + 4 \partial_r d 
\ee
as $r \rightarrow +\infty$.
The asymptotically Cartesian coordinate system we will use is that of isotropic coordinates, defined by
\be
R = r \left( 1 + \frac{r_+^6 - r_-^6}{4 r^6} \right)^{1/3} \quad,\quad
r = \frac{1}{2} \left( \sqrt{ R^6 - r_+^6 + r_-^6 } + \rho^3 \right) \,,
\ee
such that the external metric becomes
\be
ds^2 = - H^{-1}W dt^2 + f(r) d\vec{x}_8{}^2 \,,
\ee
where we can now identify $r  \equiv  |\vec{x}_8|$, and
\be
f(r) = \left( 1 + \frac{r_+^6 - r_-^6}{4 r^6} \right)^{2/3} \,.
\ee
One can now compute $B^r \approx + \frac{r_+^6-r_-^6}{r^7}$. The $r^{-7}$ is cancelled when integrating over the seven-sphere at infinity, due to the measure. Then we may define
\be
\begin{split}
M  \equiv  Q[ \xi^t,0]\Big|_\infty & = \frac{1}{16\pi G_{DFT}} \int_{t=t_0,r=\infty}  e^{-2d} \sqrt{|g|} \left(  g^{tt} g^{rr} \partial_r g_{tt} +  B^r \right) \\
 & =  \frac{1}{16\pi G_X} ( 7 r_+^6 - r_-^6) ( 2 \pi R_X ) \vol  \,,
\end{split} 
\ee
in agreement with the result of \cite{Lu:1993vt} for the ADM mass of the black string.

\subsubsection*{The first law} 

In the above we have obtained expressions for the entropy, mass and electric charge. The former two enter the first law \eqref{1stlawintegralfinal} in a simple manner. The final contribution was defined in \eqref{eleccharges}, and here gives
\be
\begin{split} 
\slashed{\delta} W & = -\frac{1}{16\pi G_{DFT}} \int_{t=t_0,R=R_0}  d^{7}x d^{2}X\: (\xi^\rho A_{ \rho}^{ M}) 
\delta\left( e^{-2d} \sqrt{|g|}  H_{MN} \mathcal{F}^{t R N } \right)\,.
\end{split} 
\ee
Since $(\xi^\rho A_{ \rho}^{ M})=A_{t z}$ is constant on the horizon, we can pull it out of the integral. We then clearly see that $\slashed{\delta} W$ is the variation of the electric charge for $A_\mu{}^M$ times the thermodynamically conjugate variable $\Psi  \equiv  A_{tz} ( R=R_0) = - 1/\alpha$, so that 
\be
\slashed{\delta} W = \Psi \delta q_{elec}  \,,
\ee
We can now put everything together to verify that the variations of the charges we calculated obey the first law of black hole thermodynamics \eqref{1stlawintegralfinal}. It is a simple calculation to indeed check that
\be
\delta M  = T \delta S + \Psi \delta q_{elec} \,,
\label{firstlawF1}
\ee
where the variations act on the parameters $r_+, r_-$.

\subsection{Non-geometric black holes?}
\label{gaufre}

We have just considered what is arguably the simplest configuration of the spacetime metric, dilaton, and Kalb-Ramond $B$-field with a horizon in the context of double field theory. In the extremal limit one obtains the fundamental string (F1) solution, which is T-dual to a pp-wave; embedding the extremal solution in double field theory  gives a pp-wave in \emph{doubled} spacetime \cite{Berkeley:2014nza}. Likewise, 1/2-BPS extremal branes in double and exceptional field theory appear as simple wave- or monopole-like configurations \cite{Berman:2014jsa, Berman:2014hna, Bakhmatov:2016kfn}. Extremal solutions are thus expected to be non-singular in any duality frame; insofar as any of them are dual to non-geometric solutions, these non-geometric duals should therefore be sensible. An example is the $5_2^2$ brane \cite{deBoer:2012ma}, which is related by two T-dualities to the NS5 brane (the magnetic dual to the F1). This and other such solutions, however, have no horizon.

What about black non-geometric solutions? Unfortunately, most known examples of exotic branes with non-geometric behaviour, such as the $5_2^2$, are of codimension 2; hence candidate ``blackened'' solutions based on these involve logarithms of the radial coordinate $R$ and thus diverge for $R\rightarrow \infty$. For this and other reasons it was argued in \cite{deBoer:2012ma} that black exotic branes should not exist.

Since the black string is most certainly not codimension 2, one could consider whether it has any sensible non-geometric duals. In the extremal case, a candidate is the electric counterpart to the $5_2^2$: this turns out to be obtained by applying Buscher dualities to the fundamental string on both the string direction $z$ and on time $t$ \cite{Bergshoeff:2011se, Sakatani:2014hba, Blair:2015eba}, and is non-geometric in the sense that it is best expressed in the bivector frame (so instead of the two-form $B_{\mu\nu}$, one has $\beta^{\mu\nu}$; see below). Indeed, it has been argued in \cite{Malek:2013sp} that this is often necessary when considering timelike dualities. 

Let us therefore dualise the black string along $t$ and $z$. The generalised metric \eqref{genlinelt} gives
\be
\begin{split} 
ds^2 & = \frac{H}{  1 - \frac{\alpha^2 ( H-1)^2}{W} } \left( - W^{-1} d\tilde t^2 + d \tilde z^2 \right) + W^{-1} dR^2 + R^2 d\Omega_7^2 \,\\
B_{tz} & = \frac{\alpha H (1-H)}{ -W + \alpha^2 (H-1)^2 } \,,\\
e^{2(\phi-\phi_0)} & = \frac{H}{| W - \alpha^2 ( H-1)^2| } \,,
\end{split}
\ee
which in fact takes the quite simple form 
\be
\begin{split} 
ds^2 & = \tilde H^{-1} \left( - d \tilde t^2 + W d \tilde z^2 \right) + W^{-1} dr^2 + r^2 d\Omega_7^2\,, \\
B & = \alpha^{-1} (1-\tilde H^{-1}) d\tilde t \wedge d\tilde z\,,\\
e^{-2\phi} & = |\tilde H| \,,
\end{split} 
\ee
where $\tilde H = 1 - r_+^6/R^6$. This has acquired a new singularity at $R^6 = r_+^6$ (which survives in the extremal limit). This is a result of a dualisation with respect to an isometry corresponding to a Killing vector whose norm squared vanishes at this value of $R$. One may approach this as involving first dualising the black string with respect to $z$ to obtain the solution \eqref{blackpp}, then dualising on $t$: the metric component $g_{tt}$ of \eqref{blackpp} is zero at exactly $R^6 = r_+^6$. The singularity is disturbing but it is conceivable that the string background is still admissible: from the worldsheet point of view, string winding modes could resolve the singularity, as observed already in \cite{Rocek:1991ps}. The target space perspective on this would be a smooth doubled spacetime possessing a section with singularity; partial results along those lines have recently appeared in the mathematical literature on topological T-duality \cite{Mathai:2011dw}.

This solution can be interpreted as a black ``negative string'', following the extensive discussion of (extremal) negative branes in \cite{Dijkgraaf:2016lym}. The extremal versions of negative branes are characterised by the appearance of naked singularities where their harmonic functions $\tilde H$ vanish. This marks the location of a ``bubble'' surrounding the brane, the interior of which contains an exotic version of string theory/M-theory \cite{Hull:1998vg, Hull:1998ym} with the spacetime signature flipped in the worldvolume directions. Indeed, see explicitly here that for $\tilde H < 0$ (inside the bubble), $\tilde z$ becomes the timelike coordinate (observe the function $W$ appearing in the $d\tilde z^2$ part of the metric) and $\tilde t$ a spacelike coordinate. It seems now that the original horizon at $W = 0$ is contained beyond the naked singularity at $\tilde H = 0$. 

To obtain a well-defined configuration (at least in the extremal limit), one can in DFT instead express the solution using
the bivector parametrisation.
This corresponds to choosing 
\be
\gM_{MN} = \begin{pmatrix}
\tilde g_{\mu\nu} & + \tilde g_{\mu \rho} \beta^{\rho \nu} \\
-\beta^{\mu\rho} \tilde g_{\rho\nu} & \tilde g^{\mu\nu} - \beta^{\mu\rho} \tilde g_{\rho\sigma} \beta^{\sigma\nu} 
\end{pmatrix} \,.
\label{eq:genmetbeta}
\ee
Some care must be taken when working with these fields (see for instance the discussions in \cite{Andriot:2011uh, Andriot:2013xca, Andriot:2014uda} where the spacetime theory, termed ``$\beta$-supergravity'', is developed). The transformation of the generalised metric under generalised diffeomorphisms implies that in the section $\tilde\partial^\mu = 0$ although both $\tilde g_{\mu\nu}$ and $\beta^{\mu\nu}$ transform as tensors under diffeomorphisms $\xi^\mu(x)$, under gauge transformations parametrised by $\lambda_\mu(x)$ one has the unusual transformations
\be
\delta_\lambda \tilde g_{\mu\nu} =- 2 \partial_{[\mu} \lambda_{\rho]} \tilde g_{\nu \sigma} \beta^{\rho\sigma} 
- \partial_{[\nu} \lambda_{\rho]} \tilde g_{\mu \sigma} \beta^{\rho\sigma} 
\quad , \quad
\delta_\lambda \beta^{\mu\nu} = - 2 ( \tilde g^{\mu\rho} \tilde g^{\nu\sigma} - \beta^{\mu\rho} \beta^{\nu\sigma}) \partial_{[\rho} \lambda_{\sigma]} \,.
\ee
The theory in this frame is therefore not a conventional theory of a metric coupled to the exotic bivector field $\beta^{\mu\nu}$. However, let us suppose we can treat $\tilde g_{\mu\nu}$ as giving a metric for the section with coordinates $x^\mu$, keeping in mind that the metric is actually modified under $\lambda_\mu$ gauge transformations. For the configuration \eqref{genlinelt} that we are considering, one finds\footnote{Observe that we only switch to the bivector parametrisation for the $(t,z,\tilde t,\tilde z)$ components of the generalised metric.}
\be
\begin{split} 
d\tilde s^2 & = - H W^{-1} d\tilde t^2 + H d\tilde z^2 + W^{-1} dR^2 + R^2 d\Omega_7^2 \,\\
\beta^{\tilde t \tilde z} & = \alpha ( H^{-1} - 1 ) \,,\\
e^{-2(\phi-\phi_0)} & = H^{-1} W \,.
\end{split} 
\ee
We see that in this frame, there is a singularity when $W=0$, with the dilaton blowing up there. On the other hand, the extremal solution with $W=1$ is sensible everywhere.

One can calculate the charges for this background. We focus on those defined at infinity. The current components can be calculated for instance by using the general formula \eqref{current_formula} and the expression for the generalised metric defined by \eqref{genlinelt}.  There is then a charge associated to $\Lambda_{\tilde z}$, which one can perhaps think of as an electric $Q$-flux:
\be
Q [ 0 , \Lambda_{\tilde z} ] \Big|_{\infty} = -6 r_-^6 \alpha ( 2 \pi R_{\tilde z} ) \vol \,.
\ee
In addition, we have a charge associated to $\xi^{\tilde t}$, which we would expect to define the mass:
\be
M  \equiv  Q[ \xi^{\tilde t} , 0 ]\Big|_{\infty} = - (5 r_+^6 + r_-^6) ( 2 \pi R_{\tilde z} ) \vol \,.
\ee
This mass is negative. In the extremal limit, this was observed in the context of DFT in \cite{Blair:2015eba, Park:2015bza}, although in fact the appearance of negative mass for the timelike dual of the non-extremal solution was shown long ago by Welch in \cite{Welch:1994qm} (this paper suggests that timelike duality does not necessarily always lead to negative mass, though).

If one uses the bivector parametrisation but takes $\partial_\mu = 0$ to be the section condition, so that gauge parameters depend on dual coordinates, then $\tilde g^{\mu\nu}$ (with upper indices) transforms as a metric under ``dual diffeomorphisms'' parametrised by $\lambda_\mu(\tilde x)$, while $\beta^{\mu\nu}$ transforms as a form, with gauge parameters $\xi^\mu$. Viewing then $\tilde g^{\mu\nu}$ as a metric for the space parametrised by the dual coordinates, we find that the solution is identical to the original black string. The charge should then presumably be defined using $J_{\tilde t}$, which is the original $J^t$ of the black string solution. It is not clear whether there is an unambiguous approach to defining mass of a solution when we allow timelike dualities. 

However, this negative mass is expected if we are indeed dealing with a ``negative brane'' though \cite{Dijkgraaf:2016lym}. Within the bubble, where the string theory is exotic, the negative mass object behaves as a standard positive mass object. 


To answer the question posed in this section: it is not obvious whether physically sensible black exotic brane or non-geometric black hole solutions exist. The results of \cite{Dijkgraaf:2016lym} seem to suggest that black ``negative branes'' might, ultimately, make sense, despite the singularity at the edge of the ``bubble'' of exotic spacetime signature. If exotic black branes do exist our formalism should be able to describe their entropy and thermodynamics.

\section{Conclusions}
\label{conclusions}

We provided a duality-invariant (under $O(n,n), n\leq D-2$) derivation of the first law of black hole thermodynamics \eqref{1stlawintegralfinal} with accompanying manifestly invariant definitions for mass \eqref{massdef} and entropy \eqref{entropy_def}. Momentum and winding ($B$-field) charge enter the first law on equal footing --- as one would expect on physical grounds \cite{Horne:1991cn} --- as electric charges of the generalised-vector-valued gauge field $A_\mu^{\phantom{\mu} M}$. Our entropy formula reduces to the area of the black hole horizon when i) the DFT fields are parametrised in the standard way \eqref{dft_spacetime_parametrisation} in terms of a spacetime metric and other fields and ii) the standard solution to the strong constraint ($\tilde \partial^\mu=0$) is used; otherwise, it is strictly more general than known results from the general relativity literature. The appearance of the entropy variation in the first law \eqref{1stlawintegralfinal} serves as a (partial) justification of the 
thermodynamic interpretation in this more general context.

Our arguments complete and extend those of Horowitz and Welch in \cite{Horowitz:1993wt}; in particular our use of the covariant phase space approach due to Wald et al. \cite{Lee:1990nz,Wald:1993nt,Iyer:1994ys} allowed for a \emph{derivation} of a formula for entropy \eqref{entropy_def} as the horizon area in the Einstein frame (this was an assumption in \cite{Horowitz:1993wt}). 

A technical advantage of our approach is that our results are largely --- and in the case of the differential form of the first law \eqref{1stlawdifferential}, entirely --- independent of any parametrisation for the generalised metric and dilaton. As an immediate corollary, \eqref{1stlawdifferential} automatically holds for any theory described in terms of the DFT Lagrangian, fields and gauge transformations as described in this paper; for instance one could apply this to the heterotic DFT of \cite{Hohm:2011ex}, to gauged supergravity (using a generalised Scherk-Schwarz reduction) \cite{Grana:2012rr}, or to the so-called $\beta$-supergravity \cite{Andriot:2011uh, Andriot:2013xca}. 

It would be remiss to not comment on deficiencies of our approach. The most prominent one is that our first law of black hole mechanics does not include magnetic charge contributions. This is because magnetic charge is not Noether charge and therefore fails to appear in \eqref{1stlawdifferential}, which expresses the conservation of Noether charge. There are at least two ways to fix this, neither of which is straightforward: the first is to write down and work with a magnetic ``Dual DFT'', the fundamental fields for which would include the magnetic dual to the $B$-field of DFT. This theory is only known at the linear level \cite{Bergshoeff:2016ncb}. The second way is to keep working with (electric) DFT but follow \cite{Keir:2013jga} in carefully taking into account ``edge'' contributions between local patches where the gauge fields are well--defined. This approach, therefore, seems to hinge on how and whether global issues are resolved in DFT. We note that it also appears that magnetic charge in DFT should 
be measured using integrals of the so-called generalised fluxes, as discussed in \cite{Blair:2015eba} -- these are defined in terms of a generalised vielbein rather than the generalised metric, and are in fact not invariant under generalised Lorentz transformations, so it seems unclear how one would obtain this expression using the present methods. For these reasons we leave the issue of magnetic charge for future work.

One might wonder about the other laws of black hole thermodynamics in a duality-manifest context. Known proofs of e.g. the second law in the general relativity literature employ concepts which are currently unavailable for DFT (geodesics, for instance), so we also leave them for future work.

Our results should generalise to the Ramond-Ramond \cite{Hohm:2011zr} and fermion sectors \cite{Jeon:2012hp} of type II DFT, once the complication of local $O(D,D)$ gauge symmetry is accounted for.

Needless to say, it will also be of interest to generalise to exceptional field theory (EFT) \cite{Hohm:2013pua}. The split parameterisation of the DFT fields used here provides an example of the tensor hierarchy structure and symmetries of EFT. In EFTs, one generally has an ${\rm E}_{n(n)}$-invariant $d$-dimensional external metric $g_{\mu\nu}$, a generalised metric $\mathcal{M}_{MN}$ for an $N$-dimensional internal extended space, and various gauge fields $A,B,\dots$ reminiscent of $A_\mu{}^M, B_{\mu\nu}$ of this paper. There is no generalised dilaton density, so the various fields transform as densities under generalised diffeomorphisms; in particular $g_{\mu\nu}$ transforms with weight $-2/(d-2)$. The entropy formula \eqref{entropy_def} should therefore be given by 
\be
\label{EFT_entropy_formula}
S \propto\int d^{d-2}x \int d^N X \: \sqrt{\det g_{d-2}} \,,
\ee
where $g_{d-2}$ denotes the pull-back of the external metric $g_{\mu\nu}$ to a cross-section of the horizon, defined for the EFT external metric $g_{\mu\nu}$ as it was for the DFT one in this paper. This integrand has weight 1, and so \eqref{EFT_entropy_formula} is manifestly invariant under external and internal generalised diffeomorphisms, as well as ${\rm E}_{n(n)}$ duality rotations.

A possible complication in the ${\rm E}_{7(7)}$ (and more generally, ${\rm E}_{n(n)}$ with $n$ odd) case is the absence of a true action with manifest ${\rm E}_{n(n)}$ invariance. In those dimensionalities, electric and magnetic charges lie on the same duality orbits, so that one can \emph{either} have a true action involving electric potentials only (thus breaking duality-invariance), \emph{or} maintain invariance at the cost of imposing a self-duality condition by hand after variations are taken. In the other cases, including ${\rm E}_{8(8)}$ and ${\rm E}_{6(6)}$, this is not an issue, and the contributions of the other fields to the first law could be worked out using an analysis similar to the one presented in this paper. It might be of interest to pursue the EFT origin of the entropy formulae for extremal black holes in $d=4$ and $d=5$, with the entropy being given in terms of duality invariant expressions involving the charges of the gauge fields \cite{Kallosh:1996uy, Cvetic:1996zq}.

It may also be interesting to pursue further the issue of timelike duality, which seemingly leads to solutions with negative mass and singularities in place of horizons (meaning that it seems one no longer has duality invariant notions of mass and entropy). Witten's 2d black hole \cite{Witten:1991yr} provides one particularly simple example of this phenomenon. In this case, one has access to a CFT description of the background, so it may be possible to combine a DFT analysis similar to that of this paper with a doubled worldsheet approach in order to investigate the subtle properties of timelike dualities within a doubled formalism.

\section*{Acknowledgements} 

We would like to thank Joe Keir, Emanuel Malek, and Daniel Thompson  for useful discussions, and Paul Townsend for reading this manuscript and bringing \cite{Dijkgraaf:2016lym} to our attention. ASA would like to thank the Theoretical Particle Physics group at Vrije Universiteit Brussel, and Alexander Sevrin in particular, for hospitality while part of this work was undertaken. ASA is supported by the UK Science and Technology Facilities Council (grant ST/L000385/1), Clare Hall College, Cambridge, and the Cambridge Trust. 
CB is supported in part by the Belgian Federal Science Policy Office through the Interuniversity Attraction Pole P7/37 ``Fundamental Interactions'', and in part by the ``FWO-Vlaanderen'' through the project G.0207.14N and by the Vrije Universiteit Brussel through the Strategic Research Program ``High-Energy Physics''.

\appendix

\section{Integration and Stokes' theorem}
\label{appendix_integration}
In this appendix we will give a version of Stokes' theorem in a form useful for double field theory. This involves formally maintaining $O(D,D)$ covariance throughout, although care should be taken when considering dualising along a coordinate transverse to a submanifold. This subtlety does not arise in the main text, because we explicitly break $O(D,D)$ in external coordinates. We will follow Naseer, who proves the codimension 1 case in \cite{Naseer:2015fba}.

Stokes' theorem is usually given as a relation involving differential forms. These do not seem to be relevant for double field theory, so we will instead derive a generalisation for the \emph{dual} statement, involving a contravariant antisymmetric tensor. For a codimension 2 submanifold of ordinary spacetime, we are integrating an antisymmetric rank-two tensor $Q^{m n}$ and the statement of Stokes' theorem is \cite{Carroll:2004st}:
\be
\int_C d^{D-1}y\: \sqrt{|\det g_{(D-1)}|} n_m \nabla_n Q^{m n} = \int_{\partial C}d^{D-2}y\: \sqrt{|\det g_{(D-2)}|} n_m \sigma_n Q^{m n}\,.
\ee
where $C$ is a codimension 1 submanifold of the $D$-dimensional spacetime with unit timelike normal $n_m$, its boundary $\partial_C$ is codimension 2 (in the spacetime) and has unit spacelike normal $\sigma_m$ and the metrics $g_{(D-1)}$ and $g_{(D-2)}$ are the induced metrics on $C$ and $\partial_C$ respectively.

Assuming $C$ is specified by $t=0$ where $t$ is a spacetime scalar and $\partial C$ is specified by the additional condition $ R=0$, we can recast the integrand on the right-hand side as
\be
\label{determinantmagic}
\sqrt{|\det g_{(D-2)}|} n_m \sigma_n Q^{m n}=\sqrt{|\det g|}\partial_m(t) \partial_n( R) Q^{mn}= \sqrt{|\det g|} Q^{t R}
\ee
where $g$ is now the full $D$-dimensional metric. The normalisation factors in $n_m$ and $\sigma_n$ have conspired with $\sqrt{|\det g_{(D-2)}|}$ to produce the determinant on the right-hand side; this is trivial when the metric is block-diagonal, and we can always put the metric in that form locally (using e.g. Gaussian normal coordinates iteratively).

The last expression is more natural for double field theory because the only integration measure readily available is the (exponential of) the generalised dilaton,
\be
e^{-2d}= \sqrt{|\det g|} e^{-2\phi}\,,
\ee
which involves the full $D$-dimensional metric. Using this we can write down Stokes' theorem for double field theory
\be
 \int_C d^{2(D-1)}X \: \partial_M \left( e^{-2  d} Q^{M N}\right) N^t_N =\int_{\partial C}  d^{2(D-2)}X\: e^{-2  d} Q^{M N} N^t_N N^ R_M
\ee
where the generalised normal vectors are $N^t_M= \partial_M (t)$, $N^ R_M=\partial_M ( R)$, $Q^{MN}$ is an antisymmetric generalised tensor of weight zero and $C$ and $\partial C$ are specified by the vanishing of the scalars $t$ and $ R$ respectively. The integrations are over the physical $(D-1)$ and $(D-2)$ dimensional submanifolds $C$ and $\partial C$ selected by the solution to the section condition. In the main text we use the more compact notation $\varepsilon_M  \equiv   N^t_M\,,\varepsilon_{MN} \equiv  N^t_{[M} N^R_{N]}$
\be
 \boxed{\int_C d^{2(D-1)}X \: \partial_M \left( e^{-2  d} Q^{M N}\right) \varepsilon_N =\int_{\partial C}  d^{2(D-2)}X\: e^{-2  d} Q^{M N} \varepsilon_{MN}\,.}
\ee
We emphasise that the epsilons thus defined are \emph{field-independent} and in fact take fixed numerical values in $(t,R)$ coordinates; in particular $\varepsilon_{t R}=-\varepsilon_{R t}=1/2$ since we are antisymmetrising with weight 1.

Stokes' theorem is trivial in ``adapted'' coordinates (where $t$ and $ R$ are part of the definition of the coordinate chart) so the only thing we have to do is verify that both integrands transform as generalised densities. This is manifest for the one on the right-hand side. For the term on the left we rewrite (dropping the superscript $t$ on the normal)
\be
\partial_P(e^{-2d} Q^{PM}) N_M= \partial_P(e^{-2d} Q^{PM} N_M)\,.
\ee
where we dropped the term involving $\partial_P N_M$ because partial derivatives commute and $Q^{MN}$ is antisymmetric. Now we only have to check that
\be
\partial_P(e^{-2d} J^P) \qquad (J^P  \equiv  Q^{PM} N_M)
\ee
transforms nicely. This is an easy calculation using the strong constraint and commuting partials:
\begin{align}
\delta_\Lambda \partial_P(e^{-2d} J^P)&=\partial_P\left[ \partial_R(\Lambda^R e^{-2d}) J^P + e^{-2d} (\Lambda^R \partial_R J^P- \partial_R \Lambda^P J^R)\right]\\ &=
\partial_P \partial_R (e^{-2d} J^P \Lambda^R)-\partial_P (e^{-2d} J^R \partial_R \Lambda^P)\\
&= \partial_R (\Lambda^R \partial_P(e^{-2d} J^P)) + \partial_R (e^{-2d} J^P \partial_P \Lambda^R)-\partial_P (e^{-2d} J^R \partial_R \Lambda^P)\\
&= \partial_R (\Lambda^R \partial_P(e^{-2d} J^P))\,,
\end{align}
which is the correct result for a generalised tensor density.

To complete the argument we note that under \emph{finite} gauge transformations, such a generalised density transforms with a Jacobian factor (see section (2.2) of \cite{Hohm:2012gk}) and cancels against the measure $d^{2(D-n)}X$ so that its integral is indeed invariant. The above argument was adapted from \cite{Naseer:2015fba}, where Stokes' theorem for the codimension 1 case can also be found. The same argument implies a Stokes' theorem for submanifolds $C$ of arbitrary higher codimension $n-1$:
\be
 \boxed{\int_C d^{2(D-(n-1))}X \: \partial_M \left( e^{-2  d} Q^{M N_1 N_2\dots N_{n-1}}\right) \varepsilon_{N_1 N_2\dots N_{n-1}} =\int_{\partial C}  d^{2(D-n)}X\: e^{-2  d} Q^{N_1 N_2\dots N_n} \varepsilon_{N_1 N_2\dots N_n}\,.}
\ee

\section{Further details of the split decomposition of DFT}

\subsection{Decomposition}

Let us relate the decomposition of the generalised metric \eqref{KK_parametrisation} to the corresponding decomposition of the spacetime fields, assuming that we have parametrised the full generalised metric in the usual manner as:
\be
H_{\hat M \hat N} = \begin{pmatrix} G - B G^{-1}B & B G^{-1} \\ - G^{-1} B & G^{-1} \end{pmatrix} \,.
\ee
Then splitting the $D$-dimensional spacetime index into a $d$-dimensional external index $\mu$ and an $n$-dimensional internal index $m$, one has
\be
\begin{split}
G_{\mu \nu} &  = g_{\mu \nu} + A_\mu{}^p A_\nu{}^q \phi_{pq} \\ 
G_{\mu m} & = A_\mu{}^p \phi_{pm} \\
G_{mn} & = \phi_{mn} 
\end{split} 
\qquad 
\begin{split} 
G^{\mu \nu} &  = g^{\mu \nu}  \\ 
G^{\mu m} & = - g^{\mu \rho} A_\rho{}^m  \\
G^{mn} & = \phi^{mn} + A_\rho{}^m A_\sigma{}^n g^{\rho \sigma} 
\end{split} 
\qquad
\begin{split}
B_{mn} & = b_{mn} \\
B_{\mu m} & = A_{\mu m} + A_{\mu}{}^p b_{pm} \\
B_{\mu \nu} & = b_{\mu \nu} -  A_{[\mu}{}^p A_{\nu] p} + A_\mu{}^p A_\nu{}^q b_{pq} 
\end{split}
\label{split_to_spacetime}
\ee
One then forms the $O(n,n)$ generalised metric out of $\phi_{mn}$ and $b_{mn}$, while $A_\mu{}^M$ has the components $A_\mu{}^i$ and $A_{\mu i}$. 

\subsection{Symmetries} 
\label{splittransfs} 

We summarise here the transformation rules of the fields in the split parametrisation. For further details, we refer the reader to \cite{Hohm:2013nja} (note that the sign of the $B$-field differs in our conventions to the one used there).

We split the $O(D,D)$ generalised diffeomorphism parameter $\Lambda^{\hat M}$ into an external diffeomorphism, external $B$-field gauge transformation and internal $O(n,n)$ generalised diffeomorphism as in the main text:
\be
\label{gauge_parameter_split}
\Lambda^{\hat M}=(\xi^\mu, \lambda_\mu, \Lambda^M)\,.
\ee
Under $O(n,n)$ generalised diffeomorphisms, the generalised metric $H_{MN}$ and dilaton $e^{-2d}$ transform in the usual manner as given in \eqref{eq:gendiffeo}. The external metric $g_{\mu\nu}$ transforms as a scalar under generalised diffeomorphisms, while one has
\be
\begin{split} 
\delta_\Lambda A_\mu{}^M & = \partial_\mu \Lambda^M + \mathcal{L}_\Lambda A_\mu{}^M \,, \\
\delta_\Lambda B_{\mu\nu} & = \Lambda^N \partial_N B_{\mu\nu} - 2 \partial_{[\mu} \Lambda^N A_{\nu]N}.
\end{split}
\ee
Meanwhile, under $\lambda_\mu$ gauge transformations, one has
\be
\begin{split}
\delta_\lambda B_{\mu\nu} & = 2 \partial_{[\mu} \lambda_{\nu]} + \partial_M \lambda_{[\mu} A_{\nu]}{}^M \,, \\
\delta_\lambda A_\mu{}^M & = - \partial^M \lambda_\mu \,.
\end{split} 
\label{eq:deltalambda} 
\ee
Finally, one has external diffeomorphisms parametrised by $\xi^\mu$, which are found to be given by
\be
\begin{split}
\delta_\xi g_{\mu \nu} & = \underline{L_\xi} g_{\mu\nu} + \delta_{\Lambda=\xi^\rho A_\rho} g_{\mu\nu} \,, \\ 
\delta_\xi A_\mu{}^M & = \xi^\nu \mathcal{F}_{\nu\mu}{}^M + H^{MN} g_{\mu\nu} \partial_N \xi^\nu+  \delta_{\Lambda=\xi^\rho A_\rho} A_\mu{}^M  + \delta_{\lambda_\sigma=-\xi^\rho C_{\sigma\rho}} A_\mu{}^M \,, \\
\delta_\xi B_{\mu \nu} & = \xi^\rho \mathcal{H}_{\mu\nu\rho} - A_{[\mu}{}^N \delta_\xi A_{\nu]  N} + \delta_{\Lambda=\xi^\rho A_\rho} B_{\mu\nu} + \delta_{\lambda_\sigma=-\xi^\rho C_{\sigma\rho}} B_{\mu\nu}\,,\\
\delta_\xi H_{MN} & =  \underline{L_\xi} H_{MN} + \delta_{\Lambda=\xi^\rho A_\rho} H_{MN}  \,,
\end{split} 
\label{eq:extdiffeos} 
\ee
where $\underline{L_\xi}$ takes the form of the conventional Lie derivative, but with $D_\mu$ in place of $\partial_\mu$. We have organised the infinitesimal gauge transformations here into covariantised gauge transformations plus terms that take the form of field-dependent gauge transformations. As is usual in EFT, the latter can be dropped when formulating the action of the theory. However, we emphasise that \eqref{eq:extdiffeos} \emph {is precisely what one gets from splitting the action of the $O(D,D)$ generalised diffeomorphism with parameter $\Lambda^{\hat M}=(\xi^\mu, 0, 0)$;} one could get rid of the field-dependent gauge transformations by considering instead $\Lambda^{\hat M}=(\xi^\mu, -\xi^\rho C_{\mu\rho} , \xi^\rho A_\rho{}^M )$, but as this introduces an explicit field-dependence in the gauge transformation parameter we do not do this.

\subsection{Other components of the current} 
\label{Japp} 

The remaining components of the current, which do not contribute to the charges in our set-up are:
\be
\begin{split}
e^{2\hat d} J^{\mu M} & = 
H^{MN} \partial_N \xi^\mu - g^{\mu \nu} \underline{\nabla}_\nu \tilde \Lambda^M 
- \eta^{MN} \partial_N ( g^{\mu \nu} \tilde \lambda_\nu ) \\ & 
- \xi^\nu H^{MN} g_{\nu \rho} \partial_N g^{\mu \rho} 
+ \tilde \Lambda^N g^{\mu \nu} H_{NP} D_\nu H^{MP} \\ & 
+ \tilde \lambda_\nu H^M{}_P \mathcal{F}^{\mu \nu P} 
- A_\lambda{}^M J^{\mu \lambda} e^{2d }\,,
\end{split} 
\ee
\be
\begin{split}
e^{2\hat d} J^{MN} & = 
2 \partial_Q \tilde \Lambda^{[M} H^{N]Q} 
- 2 \partial_Q ( \tilde \Lambda^P H_P{}^{[M} ) \eta^{N]Q} 
\\ & 
+ 2 \tilde \Lambda^P H_{PQ} H^{K[M} \partial_K H^{N]Q} 
- \tilde \Lambda^P H^K{}_P H^{[M}{}_Q \partial_K H^{N]Q} 
\\ &
- \tilde \lambda_\mu g^{\mu \nu} \eta_{PQ} H^{P[M} D_\nu H^{N]Q} 
\\ & 
- 2 A_\mu{}^{[M|} J^{\mu |N]} e^{2d} 
- A_\mu{}^M A_\nu{}^N J^{\mu \nu} e^{2d} \,.
\end{split}
\ee
The remaining components of the boundary vector are:
\be
B^M=-\partial_N H^{MN} + 4 H^{MN} \partial_N d
+ D_\mu ( g^{\mu\nu} A_\nu{}^M) - \partial_N A_\nu{}^N g^{\mu\nu} A_\mu{}^M 
- g^{\mu\nu} A_\nu{}^M D_\mu ( 4 d - \ln g ) \,,
\ee
and
\be
\begin{split}
B_\mu & = 
+ D_\mu ( g^{\nu\rho} C_{\mu\rho}) - g^{\nu\rho} C_{\mu\rho} D_\nu ( 4d -\ln g ) 
- \partial_N A_\nu{}^N g^{\nu\rho} C_{\mu\rho} 
\\ & - \partial_N ( H^M{}_P A_\mu{}^P ) + H^M{}_P A_\mu{}^P \partial_M ( 4d - \ln g) \,.
\end{split}
\ee

\bibliography{EvenNewerBib}

\providecommand{\href}[2]{#2}\begingroup\raggedright\begin{thebibliography}{10}

\bibitem{Callan:1985ia}
J.~Callan, Curtis~G., E.~Martinec, M.~Perry, and D.~Friedan, {\it {Strings in
  Background Fields}},  {\em Nucl.Phys.} {\bf B262} (1985) 593.

\bibitem{Buscher:1987qj}
T.~Buscher, {\it {Path Integral Derivation of Quantum Duality in Nonlinear
  Sigma Models}},  {\em Phys.Lett.} {\bf B201} (1988) 466.

\bibitem{Buscher:1987sk}
T.~Buscher, {\it {A Symmetry of the String Background Field Equations}},  {\em
  Phys.Lett.} {\bf B194} (1987) 59.

\bibitem{Rocek:1991ps}
M.~Rocek and E.~P. Verlinde, {\it {Duality, quotients, and currents}},  {\em
  Nucl.Phys.} {\bf B373} (1992) 630--646,
  [\href{http://xxx.lanl.gov/abs/hep-th/9110053}{{\tt hep-th/9110053}}].

\bibitem{Duff:1989tf}
M.~Duff, {\it {Duality rotations in string theory}},  {\em Nucl.Phys.} {\bf
  B335} (1990) 610.

\bibitem{Siegel:1993th}
W.~Siegel, {\it {Superspace duality in low-energy superstrings}},  {\em
  Phys.Rev.} {\bf D48} (1993) 2826--2837,
  [\href{http://xxx.lanl.gov/abs/hep-th/9305073}{{\tt hep-th/9305073}}].

\bibitem{Siegel:1993xq}
W.~Siegel, {\it {Two vierbein formalism for string inspired axionic gravity}},
  {\em Phys.Rev.} {\bf D47} (1993) 5453--5459,
  [\href{http://xxx.lanl.gov/abs/hep-th/9302036}{{\tt hep-th/9302036}}].

\bibitem{Hull:2009mi}
C.~Hull and B.~Zwiebach, {\it {Double Field Theory}},  {\em JHEP} {\bf 0909}
  (2009) 099, [\href{http://xxx.lanl.gov/abs/0904.4664}{{\tt
  arXiv:0904.4664}}].

\bibitem{Hull:2009zb}
C.~Hull and B.~Zwiebach, {\it {The Gauge algebra of double field theory and
  Courant brackets}},  {\em JHEP} {\bf 0909} (2009) 090,
  [\href{http://xxx.lanl.gov/abs/0908.1792}{{\tt arXiv:0908.1792}}].

\bibitem{Hohm:2010jy}
O.~Hohm, C.~Hull, and B.~Zwiebach, {\it {Background independent action for
  double field theory}},  {\em JHEP} {\bf 1007} (2010) 016,
  [\href{http://xxx.lanl.gov/abs/1003.5027}{{\tt arXiv:1003.5027}}].

\bibitem{Hohm:2010pp}
O.~Hohm, C.~Hull, and B.~Zwiebach, {\it {Generalized metric formulation of
  double field theory}},  {\em JHEP} {\bf 1008} (2010) 008,
  [\href{http://xxx.lanl.gov/abs/1006.4823}{{\tt arXiv:1006.4823}}].

\bibitem{Aldazabal:2013sca}
G.~Aldazabal, D.~Marqu{\'e}s, and C.~N{\'u}{\~n}ez, {\it {Double Field Theory:
  A Pedagogical Review}},  {\em Class.Quant.Grav.} {\bf 30} (2013) 163001,
  [\href{http://xxx.lanl.gov/abs/1305.1907}{{\tt arXiv:1305.1907}}].

\bibitem{deBoer:2012ma}
J.~de~Boer and M.~Shigemori, {\it {Exotic Branes in String Theory}},  {\em
  Phys.Rept.} {\bf 532} (2013) 65--118,
  [\href{http://xxx.lanl.gov/abs/1209.6056}{{\tt arXiv:1209.6056}}].

\bibitem{Berkeley:2014nza}
J.~Berkeley, D.~S. Berman, and F.~J. Rudolph, {\it {Strings and Branes are
  Waves}},  {\em JHEP} {\bf 06} (2014) 006,
  [\href{http://xxx.lanl.gov/abs/1403.7198}{{\tt arXiv:1403.7198}}].

\bibitem{Berman:2014jsa}
D.~S. Berman and F.~J. Rudolph, {\it {Branes are Waves and Monopoles}},  {\em
  JHEP} {\bf 05} (2015) 015, [\href{http://xxx.lanl.gov/abs/1409.6314}{{\tt
  arXiv:1409.6314}}].

\bibitem{Berman:2014hna}
D.~S. Berman and F.~J. Rudolph, {\it {Strings, Branes and the Self-dual
  Solutions of Exceptional Field Theory}},  {\em JHEP} {\bf 05} (2015) 130,
  [\href{http://xxx.lanl.gov/abs/1412.2768}{{\tt arXiv:1412.2768}}].

\bibitem{Bakhmatov:2016kfn}
I.~Bakhmatov, A.~Kleinschmidt, and E.~T. Musaev, {\it {Non-geometric branes are
  DFT monopoles}},  \href{http://xxx.lanl.gov/abs/1607.0545}{{\tt
  arXiv:1607.0545}}.

\bibitem{Blair:2015eba}
C.~D.~A. Blair, {\it {Conserved Currents of Double Field Theory}},  {\em JHEP}
  {\bf 04} (2016) 180, [\href{http://xxx.lanl.gov/abs/1507.0754}{{\tt
  arXiv:1507.0754}}].

\bibitem{Park:2015bza}
J.-H. Park, S.-J. Rey, W.~Rim, and Y.~Sakatani, {\it {O(D, D) covariant Noether
  currents and global charges in double field theory}},  {\em JHEP} {\bf 11}
  (2015) 131, [\href{http://xxx.lanl.gov/abs/1507.0754}{{\tt
  arXiv:1507.0754}}].

\bibitem{Naseer:2015fba}
U.~Naseer, {\it {Canonical formulation and conserved charges of double field
  theory}},  {\em JHEP} {\bf 10} (2015) 158,
  [\href{http://xxx.lanl.gov/abs/1508.0084}{{\tt arXiv:1508.0084}}].

\bibitem{Mohaupt:2000mj}
T.~Mohaupt, {\it {Black hole entropy, special geometry and strings}},  {\em
  Fortsch. Phys.} {\bf 49} (2001) 3--161,
  [\href{http://xxx.lanl.gov/abs/hep-th/0007195}{{\tt hep-th/0007195}}].

\bibitem{Horowitz:1993wt}
G.~T. Horowitz and D.~L. Welch, {\it {Duality invariance of the Hawking
  temperature and entropy}},  {\em Phys. Rev.} {\bf D49} (1994) 590--594,
  [\href{http://xxx.lanl.gov/abs/hep-th/9308077}{{\tt hep-th/9308077}}].

\bibitem{Lee:1990nz}
J.~Lee and R.~M. Wald, {\it {Local symmetries and constraints}},  {\em J. Math.
  Phys.} {\bf 31} (1990) 725--743.

\bibitem{Wald:1993nt}
R.~M. Wald, {\it {Black hole entropy is the Noether charge}},  {\em Phys. Rev.}
  {\bf D48} (1993) 3427--3431,
  [\href{http://xxx.lanl.gov/abs/gr-qc/9307038}{{\tt gr-qc/9307038}}].

\bibitem{Iyer:1994ys}
V.~Iyer and R.~M. Wald, {\it {Some properties of Noether charge and a proposal
  for dynamical black hole entropy}},  {\em Phys. Rev.} {\bf D50} (1994)
  846--864, [\href{http://xxx.lanl.gov/abs/gr-qc/9403028}{{\tt
  gr-qc/9403028}}].

\bibitem{Jeon:2010rw}
I.~Jeon, K.~Lee, and J.-H. Park, {\it {Differential geometry with a projection:
  Application to double field theory}},  {\em JHEP} {\bf 1104} (2011) 014,
  [\href{http://xxx.lanl.gov/abs/1011.1324}{{\tt arXiv:1011.1324}}].

\bibitem{Jeon:2011cn}
I.~Jeon, K.~Lee, and J.-H. Park, {\it {Stringy differential geometry, beyond
  Riemann}},  {\em Phys.Rev.} {\bf D84} (2011) 044022,
  [\href{http://xxx.lanl.gov/abs/1105.6294}{{\tt arXiv:1105.6294}}].

\bibitem{Hohm:2011si}
O.~Hohm and B.~Zwiebach, {\it {On the Riemann Tensor in Double Field Theory}},
  {\em JHEP} {\bf 1205} (2012) 126,
  [\href{http://xxx.lanl.gov/abs/1112.5296}{{\tt arXiv:1112.5296}}].

\bibitem{Berman:2013uda}
D.~S. Berman, C.~D.~A. Blair, E.~Malek, and M.~J. Perry, {\it {The $O_{D,D}$
  geometry of string theory}},  {\em Int.J.Mod.Phys.} {\bf A29} (2014), no.~15
  1450080, [\href{http://xxx.lanl.gov/abs/1303.6727}{{\tt arXiv:1303.6727}}].

\bibitem{Hohm:2014xsa}
O.~Hohm and B.~Zwiebach, {\it {Double Field Theory at Order $\alpha'$}},
  \href{http://xxx.lanl.gov/abs/1407.3803}{{\tt arXiv:1407.3803}}.

\bibitem{Marques:2015vua}
D.~Marques and C.~A. Nunez, {\it {T-duality and α′-corrections}},  {\em
  JHEP} {\bf 10} (2015) 084, [\href{http://xxx.lanl.gov/abs/1507.0065}{{\tt
  arXiv:1507.0065}}].

\bibitem{Hull:2014mxa}
C.~M. Hull, {\it {Finite Gauge Transformations and Geometry in Double Field
  Theory}},  {\em JHEP} {\bf 04} (2015) 109,
  [\href{http://xxx.lanl.gov/abs/1406.7794}{{\tt arXiv:1406.7794}}].

\bibitem{Hohm:2013nja}
O.~Hohm and H.~Samtleben, {\it {Gauge theory of Kaluza-Klein and winding
  modes}},  {\em Phys.Rev.} {\bf D88} (2013) 085005,
  [\href{http://xxx.lanl.gov/abs/1307.0039}{{\tt arXiv:1307.0039}}].

\bibitem{Compere:2006my}
G.~Compere, {\it {An introduction to the mechanics of black holes}},  in {\em
  {2nd Modave Summer School in Theoretical Physics Modave, Belgium, August
  6-12, 2006}}, 2006.
\newblock \href{http://xxx.lanl.gov/abs/gr-qc/0611129}{{\tt gr-qc/0611129}}.

\bibitem{Howe:1989uk}
P.~S. Howe and P.~K. Townsend, {\it {Chern-Simons Quantum Mechanics}},  {\em
  Class. Quant. Grav.} {\bf 7} (1990) 1655--1668.

\bibitem{wald1990identically}
R.~M. Wald, {\it On identically closed forms locally constructed from a field},
   {\em Journal of mathematical physics} {\bf 31} (1990), no.~10 2378--2384.

\bibitem{Berman:2011kg}
D.~S. Berman, E.~T. Musaev, and M.~J. Perry, {\it {Boundary Terms in
  Generalized Geometry and doubled field theory}},  {\em Phys.Lett.} {\bf B706}
  (2011) 228--231, [\href{http://xxx.lanl.gov/abs/1110.3097}{{\tt
  arXiv:1110.3097}}].

\bibitem{Hohm:2013pua}
O.~Hohm and H.~Samtleben, {\it {Exceptional Form of D=11 Supergravity}},  {\em
  Phys.Rev.Lett.} {\bf 111} (2013) 231601,
  [\href{http://xxx.lanl.gov/abs/1308.1673}{{\tt arXiv:1308.1673}}].

\bibitem{Wang:2015hca}
Y.-N. Wang, {\it {Generalized Cartan Calculus in general dimension}},  {\em
  JHEP} {\bf 07} (2015) 114, [\href{http://xxx.lanl.gov/abs/1504.0478}{{\tt
  arXiv:1504.0478}}].

\bibitem{Cadabra}
K.~Peeters, {\it {A field-theory motivated approach to symbolic computer
  algebra}},  {\em Comp.Phys.Comm.} {\bf 176} (2007)
  [\href{http://xxx.lanl.gov/abs/cs/0608005}{{\tt cs/0608005}}].

\bibitem{Peeters:2007wn}
K.~Peeters, {\it {Introducing Cadabra: A Symbolic computer algebra system for
  field theory problems}},  \href{http://xxx.lanl.gov/abs/hep-th/0701238}{{\tt
  hep-th/0701238}}.

\bibitem{Racz:1992bp}
I.~Racz and R.~M. Wald, {\it {Extension of space-times with Killing horizon}},
  {\em Class. Quant. Grav.} {\bf 9} (1992) 2643--2656.

\bibitem{Mukohyama:1998mq}
S.~Mukohyama, {\it {On the Noether charge form of the first law of black hole
  mechanics}},  {\em Phys. Rev.} {\bf D59} (1999) 064009,
  [\href{http://xxx.lanl.gov/abs/gr-qc/9809050}{{\tt gr-qc/9809050}}].

\bibitem{Jacobson:1993vj}
T.~Jacobson, G.~Kang, and R.~C. Myers, {\it {On black hole entropy}},  {\em
  Phys. Rev.} {\bf D49} (1994) 6587--6598,
  [\href{http://xxx.lanl.gov/abs/gr-qc/9312023}{{\tt gr-qc/9312023}}].

\bibitem{Copsey:2005se}
K.~Copsey and G.~T. Horowitz, {\it {The Role of dipole charges in black hole
  thermodynamics}},  {\em Phys. Rev.} {\bf D73} (2006) 024015,
  [\href{http://xxx.lanl.gov/abs/hep-th/0505278}{{\tt hep-th/0505278}}].

\bibitem{Keir:2013jga}
J.~Keir, {\it {Stability, Instability, Canonical Energy and Charged Black
  Holes}},  {\em Class. Quant. Grav.} {\bf 31} (2014), no.~3 035014,
  [\href{http://xxx.lanl.gov/abs/1306.6087}{{\tt arXiv:1306.6087}}].

\bibitem{Horowitz:1991cd}
G.~T. Horowitz and A.~Strominger, {\it {Black strings and P-branes}},  {\em
  Nucl. Phys.} {\bf B360} (1991) 197--209.

\bibitem{Horne:1991cn}
J.~H. Horne, G.~T. Horowitz, and A.~R. Steif, {\it {An Equivalence between
  momentum and charge in string theory}},  {\em Phys. Rev. Lett.} {\bf 68}
  (1992) 568--571, [\href{http://xxx.lanl.gov/abs/hep-th/9110065}{{\tt
  hep-th/9110065}}].

\bibitem{Lu:1993vt}
J.~X. Lu, {\it {ADM masses for black strings and p-branes}},  {\em Phys. Lett.}
  {\bf B313} (1993) 29--34, [\href{http://xxx.lanl.gov/abs/hep-th/9304159}{{\tt
  hep-th/9304159}}].

\bibitem{Bergshoeff:2011se}
E.~A. Bergshoeff, T.~Ort{\'i}n, and F.~Riccioni, {\it {Defect Branes}},  {\em
  Nucl.Phys.} {\bf B856} (2012) 210--227,
  [\href{http://xxx.lanl.gov/abs/1109.4484}{{\tt arXiv:1109.4484}}].

\bibitem{Sakatani:2014hba}
Y.~Sakatani, {\it {Exotic branes and non-geometric fluxes}},  {\em JHEP} {\bf
  03} (2015) 135, [\href{http://xxx.lanl.gov/abs/1412.8769}{{\tt
  arXiv:1412.8769}}].

\bibitem{Malek:2013sp}
E.~Malek, {\it {Timelike U-dualities in Generalised Geometry}},  {\em JHEP}
  {\bf 1311} (2013) 185, [\href{http://xxx.lanl.gov/abs/1301.0543}{{\tt
  arXiv:1301.0543}}].

\bibitem{Mathai:2011dw}
V.~Mathai and S.~Wu, {\it {Topology and Flux of T-Dual Manifolds with Circle
  Actions}},  {\em Commun. Math. Phys.} {\bf 316} (2012) 279--286,
  [\href{http://xxx.lanl.gov/abs/1108.5045}{{\tt arXiv:1108.5045}}].

\bibitem{Dijkgraaf:2016lym}
R.~Dijkgraaf, B.~Heidenreich, P.~Jefferson, and C.~Vafa, {\it {Negative Branes,
  Supergroups and the Signature of Spacetime}},
  \href{http://xxx.lanl.gov/abs/1603.0566}{{\tt arXiv:1603.0566}}.

\bibitem{Hull:1998vg}
C.~M. Hull, {\it {Timelike T duality, de Sitter space, large N gauge theories
  and topological field theory}},  {\em JHEP} {\bf 9807} (1998) 021,
  [\href{http://xxx.lanl.gov/abs/hep-th/9806146}{{\tt hep-th/9806146}}].

\bibitem{Hull:1998ym}
C.~M. Hull, {\it {Duality and the signature of space-time}},  {\em JHEP} {\bf
  9811} (1998) 017, [\href{http://xxx.lanl.gov/abs/hep-th/9807127}{{\tt
  hep-th/9807127}}].

\bibitem{Andriot:2011uh}
D.~Andriot, M.~Larfors, D.~L{\"u}st, and P.~Patalong, {\it {A ten-dimensional
  action for non-geometric fluxes}},  {\em JHEP} {\bf 1109} (2011) 134,
  [\href{http://xxx.lanl.gov/abs/1106.4015}{{\tt arXiv:1106.4015}}].

\bibitem{Andriot:2013xca}
D.~Andriot and A.~Betz, {\it {$\beta$-supergravity: a ten-dimensional theory
  with non-geometric fluxes, and its geometric framework}},  {\em JHEP} {\bf
  1312} (2013) 083, [\href{http://xxx.lanl.gov/abs/1306.4381}{{\tt
  arXiv:1306.4381}}].

\bibitem{Andriot:2014uda}
D.~Andriot and A.~Betz, {\it {NS-branes, source corrected Bianchi identities,
  and more on backgrounds with non-geometric fluxes}},  {\em JHEP} {\bf 1407}
  (2014) 059, [\href{http://xxx.lanl.gov/abs/1402.5972}{{\tt
  arXiv:1402.5972}}].

\bibitem{Welch:1994qm}
D.~L. Welch, {\it {Timelike duality}},  {\em Phys. Rev.} {\bf D50} (1994)
  6404--6411, [\href{http://xxx.lanl.gov/abs/hep-th/9405070}{{\tt
  hep-th/9405070}}].

\bibitem{Hohm:2011ex}
O.~Hohm and S.~K. Kwak, {\it {Double Field Theory Formulation of Heterotic
  Strings}},  {\em JHEP} {\bf 1106} (2011) 096,
  [\href{http://xxx.lanl.gov/abs/1103.2136}{{\tt arXiv:1103.2136}}].

\bibitem{Grana:2012rr}
M.~Gra{\~n}a and D.~Marqu{\'e}s, {\it {Gauged Double Field Theory}},  {\em
  JHEP} {\bf 1204} (2012) 020, [\href{http://xxx.lanl.gov/abs/1201.2924}{{\tt
  arXiv:1201.2924}}].

\bibitem{Bergshoeff:2016ncb}
E.~A. Bergshoeff, O.~Hohm, V.~A. Penas, and F.~Riccioni, {\it {Dual Double
  Field Theory}},  {\em JHEP} {\bf 06} (2016) 026,
  [\href{http://xxx.lanl.gov/abs/1603.0738}{{\tt arXiv:1603.0738}}].

\bibitem{Hohm:2011zr}
O.~Hohm, S.~K. Kwak, and B.~Zwiebach, {\it {Unification of Type II Strings and
  T-duality}},  {\em Phys.Rev.Lett.} {\bf 107} (2011) 171603,
  [\href{http://xxx.lanl.gov/abs/1106.5452}{{\tt arXiv:1106.5452}}].

\bibitem{Jeon:2012hp}
I.~Jeon, K.~Lee, J.-H. Park, and Y.~Suh, {\it {Stringy Unification of Type IIA
  and IIB Supergravities under N=2 D=10 Supersymmetric Double Field Theory}},
  {\em Phys.Lett.} {\bf B723} (2013) 245--250,
  [\href{http://xxx.lanl.gov/abs/1210.5078}{{\tt arXiv:1210.5078}}].

\bibitem{Kallosh:1996uy}
R.~Kallosh and B.~Kol, {\it {E(7) symmetric area of the black hole horizon}},
  {\em Phys. Rev.} {\bf D53} (1996) R5344--R5348,
  [\href{http://xxx.lanl.gov/abs/hep-th/9602014}{{\tt hep-th/9602014}}].

\bibitem{Cvetic:1996zq}
M.~Cvetic and C.~M. Hull, {\it {Black holes and U duality}},  {\em Nucl. Phys.}
  {\bf B480} (1996) 296--316,
  [\href{http://xxx.lanl.gov/abs/hep-th/9606193}{{\tt hep-th/9606193}}].

\bibitem{Witten:1991yr}
E.~Witten, {\it {On string theory and black holes}},  {\em Phys. Rev.} {\bf
  D44} (1991) 314--324.

\bibitem{Carroll:2004st}
S.~M. Carroll, {\em {Spacetime and geometry: An introduction to general
  relativity}}.
\newblock 2004.

\bibitem{Hohm:2012gk}
O.~Hohm and B.~Zwiebach, {\it {Large Gauge Transformations in Double Field
  Theory}},  {\em JHEP} {\bf 1302} (2013) 075,
  [\href{http://xxx.lanl.gov/abs/1207.4198}{{\tt arXiv:1207.4198}}].

\end{thebibliography}\endgroup
\bibliographystyle{JHEP}

\end{document}